\documentclass[11pt,a4paper,pdfa=true]{article}
\pdfoutput=1
\usepackage{jcappub}
\usepackage{color}
\usepackage[normalem]{ulem}

\usepackage[T1]{fontenc}
\usepackage{graphicx,amssymb,amsmath,wasysym}

\usepackage{a4wide}

\usepackage{youngtab}
\usepackage[english]{babel} 
\usepackage{amsmath}
\usepackage{float}
\usepackage{amssymb}
\usepackage{braket}
\usepackage{amsfonts,bbm}
\usepackage{mathrsfs}
\usepackage{amsfonts}
\usepackage{dsfont,tensor}
\usepackage{enumerate}
\usepackage{graphicx}
\usepackage[table]{xcolor}
\usepackage{bbm}
\usepackage{array}
\usepackage{booktabs}
\usepackage{units}
\usepackage{textcomp}
\usepackage{mathtools}
\usepackage{cancel}
\usepackage{slashed}
\usepackage{breqn}
\usepackage{slashed}
\newcommand{\dd}{\text{d}}

\newcommand{\mx}{m_{\chi}}
\newcommand{\tx}{T_{\chi}}
\newcommand{\vx}{v_{\chi}}

\newcommand{\pts}[1]{\phantom{.}\hfill(\textit{#1~point}\ifthenelse{\equal{#1}{1}}{}{\textit{s}})}

\newcommand{\Msun}{{\ifmmode{{\rm
        {M_{\odot}}}}\else{${\rm{M_{\odot}}}$}\fi}}

\begin{document}


\vspace{2cm}

\hfill {\it In memory of Haim Goldberg}
	
\title{Dark matter in the Sun: \\ scattering off electrons vs nucleons}

\author[1]{Raghuveer Garani}
\author[2]{and Sergio Palomares-Ruiz}

\affiliation[1]{Bethe Center for Theoretical Physics and Physikalisches Institut, Universit\"at Bonn, Nu\ss allee 12, D-53115 Bonn, Germany}
\affiliation[2]{Instituto de F\'isica Corpuscular (IFIC), CSIC-Universitat de Val\`encia, \\  
	Apartado de Correos 22085, E-46071 Val\`encia, Spain} 

\emailAdd{garani@th.physik.uni-bonn.de, sergiopr@ific.uv.es}

\abstract{The annihilation of dark matter (DM) particles accumulated in the Sun could produce a flux of neutrinos, which is potentially detectable with neutrino detectors/telescopes and the DM elastic scattering cross section can be constrained. Although the process of DM capture in astrophysical objects like the Sun is commonly assumed to be due to interactions only with nucleons, there are scenarios in which tree-level DM couplings to quarks are absent, and even if loop-induced interactions with nucleons are allowed, scatterings off electrons could be the dominant capture mechanism. We consider this possibility and study in detail all the ingredients necessary to compute the neutrino production rates from DM annihilations in the Sun (capture, annihilation and evaporation rates) for velocity-independent and isotropic, velocity-dependent and isotropic and momentum-dependent scattering cross sections for DM interactions with electrons and compare them with the results obtained for the case of interactions with nucleons. Moreover, we improve the usual calculations in a number of ways and provide analytical expressions in three appendices. Interestingly, we find that the evaporation mass in the case of interactions with electrons could be below the GeV range, depending on the high-velocity tail of the DM distribution in the Sun, which would open a new mass window for searching for this type of scenarios.}

\maketitle

\section{Introduction}
\label{sec:intro}

Dark matter (DM) particles in the galactic halo could be brought into close orbits around the Sun after scattering off solar nuclei. Subsequent scatterings could finally capture those DM particles inside the Sun and thermalize them. It has been three decades since the effects of DM particles accumulated in the Sun were originally considered to solve the {\it solar neutrino problem} by modifying the energy transfer in the Sun~\cite{Steigman:1997vs, Spergel:1984re, Faulkner:1985rm}. However, these first papers did not attempt to explain the physical origin of the required DM concentration, i.e., how solar capture of galactic DM particles would proceed, which was studied for the first time in Ref.~\cite{Press:1985ug} (and later refined in Refs.~\cite{Gould:1987ju, Gould:1987ir}).  Soon after those seminal works, it was realized that annihilations of DM particles accumulated in the Sun would give rise to a neutrino flux, potentially detectable at neutrino detectors~\cite{Silk:1985ax, Krauss:1985aaa, Freese:1985qw, Hagelin:1986gv, Gaisser:1986ha, Srednicki:1986vj, Griest:1986yu}. Since then, this is one of the existing strategies to indirectly detect DM, which is in turn complementary to DM direct searches, given that in both cases the signal would be proportional to the DM elastic scattering cross section.  Indeed, numerous studies have evaluated the prospects of detection of the potential high-energy neutrino flux~\cite{Kamionkowski:1991nj, Bottino:1991dy, Halzen:1991kh, Gandhi:1993ce, Bottino:1994xp, Bergstrom:1996kp, Bergstrom:1998xh, Barger:2001ur, Bertin:2002ky, Hooper:2002gs, Bueno:2004dv,  Cirelli:2005gh, Halzen:2005ar, Mena:2007ty, Lehnert:2007fv, Barger:2007xf, Barger:2007hj, Blennow:2007tw, Liu:2008kz, Hooper:2008cf, Wikstrom:2009kw, Nussinov:2009ft, Menon:2009qj, Buckley:2009kw, Zentner:2009is, Ellis:2009ka, Esmaili:2009ks, Ellis:2011af, Bell:2011sn, Kappl:2011kz,  Agarwalla:2011yy, Chen:2011vda, Kundu:2011ek, Rott:2011fh, Das:2011yr, Kumar:2012uh, Bell:2012dk, Silverwood:2012tp, Blennow:2013pya, Arina:2013jya, Liang:2013dsa, Ibarra:2013eba, Albuquerque:2013xna, Baratella:2013fya, Guo:2013ypa, Ibarra:2014vya, Chen:2014oaa, Blumenthal:2014cwa, Catena:2015iea, Chen:2015uha, Belanger:2015hra, Heisig:2015ira, Danninger:2014xza, Blennow:2015hzp, Murase:2016nwx, Lopes:2016ezf, Baum:2016oow, Allahverdi:2016fvl} and of that of neutrinos in the $\cal{O}$(10-100)~MeV range~\cite{Rott:2012qb, Bernal:2012qh, Rott:2015nma, Rott:2016mzs} using neutrino detectors/telescopes~\cite{Desai:2004pq, Desai:2007ra, Abbasi:2009uz, Abbasi:2009vg, Tanaka:2011uf, IceCube:2011aj, Scott:2012mq, Aartsen:2012kia, Adrian-Martinez:2013ayv, Avrorin:2014swy, Choi:2015ara, Aartsen:2016exj, Adrian-Martinez:2016ujo, Aartsen:2016zhm}. 

All these works have focused on DM-nucleon interactions. However, the possibility of DM particles having no direct couplings to quarks, but only to leptons, the so-called leptophilic scenarios, has been extensively considered in the literature; to alleviate the conflict of the DM interpretation~\cite{Bernabei:2007gr, Dedes:2009bk, Kopp:2009et, Feldstein:2010su, Chang:2014tea, Bell:2014tta, Foot:2014xwa, Roberts:2016xfw} between the signal observed at the DAMA experiment~\cite{Bernabei:2013xsa} and the null results of other direct searches~\cite{Felizardo:2011uw, Abe:2015eos, Agnese:2015nto, Amole:2016pye, Aprile:2016wwo, Tan:2016zwf, Akerib:2016vxi, Aprile:2016swn, Aprile:2017yea}, for future strategies to search for sub-GeV DM particles with direct detection experiments~\cite{Essig:2011nj, Chen:2015pha, Lee:2015qva, Essig:2015cda}, within the context of cosmic-ray anomalies in order to explain the positron, but not antiproton, excess~\cite{Fox:2008kb, Cao:2009yy, Bi:2009uj, Ibarra:2009bm, Cohen:2009fz, Cavasonza:2016qem} seen by different experiments~\cite{Adriani:2008zr, Chang:2008aa, Adriani:2010rc, FermiLAT:2011ab, Accardo:2014lma, Aguilar:2016kjl}, to reduce the tension of the observed anomaly in the muon magnetic moment~\cite{Agrawal:2014ufa} or as potential signals in collider searches~\cite{DEramo:2017zqw}. 

Even in the case of tree-level DM couplings to electrons, in general, loop-induced DM-quark couplings are also present by coupling photons to virtual leptons~\cite{Kopp:2009et}. Therefore, DM would be captured in the Sun by both, interactions off solar electrons via tree-level processes and interactions off solar nuclei via loop processes. However, there are cases in which no loop-induced DM-quark contribution is present, such as axial vector couplings and thus, only DM capture by electrons is possible. Neutrino signals for leptophilic scenarios have been considered in Ref.~\cite{Kopp:2009et}. In that work, a constant (velocity-independent and isotropic) cross section was assumed to compute the solar capture rate of DM particles. However, DM-electron (and DM-nucleon) interactions could have a more complicated structure and non-trivial dependencies on the relative velocity ($v_{\rm rel}$) and the scattering angle ($\theta_{\rm cm}$) do appear for various operators~\cite{Fan:2010gt, Fitzpatrick:2012ix, Anand:2013yka, Hill:2013hoa, Gresham:2014vja, Panci:2014gga, Catena:2014epa, Gluscevic:2014vga, Catena:2014hla, Gluscevic:2015sqa, Dent:2015zpa, Catena:2015vpa, Kavanagh:2015jma, Catena:2015uha, Gazda:2016mrp}. Indeed, these possibilities, assuming couplings only to quarks, have been recently considered in the context of high-energy neutrino signals from the Sun~\cite{Kumar:2012uh, Guo:2013ypa, Liang:2013dsa}, to reduce the tension between solar models and helioseismological data~\cite{Lopes:2013xua, Vincent:2013lua, Lopes:2014aoa, Vincent:2014jia, Vincent:2015gqa, Vincent:2016dcp}, and to allow for a better compatibility among different results from direct searches~\cite{Masso:2009mu, Chang:2009yt, Chang:2010en, Barger:2010gv, Fitzpatrick:2010br, Foot:2011pi, Schwetz:2011xm, Farina:2011pw, Fornengo:2011sz, DelNobile:2012tx, Foot:2012cs, Fitzpatrick:2012ib, Catena:2014uqa, Barello:2014uda, Catena:2015uua, Catena:2016hoj, Rogers:2016jrx}.

In this work, we present general results for the solar DM capture, annihilation and evaporation rates, as well as for the resulting neutrino fluxes from DM annihilations at production, for the cases of interactions with electrons with constant, $v_{\rm rel}^2$-dependent and transfer momentum ($q^2$)-dependent elastic scattering cross sections. All our results are compared to those obtained for the case of DM interactions with nucleons. We perform all computations taking into account thermal effects and study their importance. Moreover, we improve over the common calculation of the rates in a number of ways. We consistently compute the temperature in the regime of weak cross sections (Knudsen limit or optically thin regime) for each case including the effect of evaporation and the truncation of the DM velocity distribution, for which we also consider several cutoff velocities. We compute the minimum DM mass for which evaporation is not efficient enough to reduce the number of captured DM particles and find that, for the case of DM-electron scatterings, depending on the cutoff velocity, the minimum testable mass could be significantly smaller (below GeV) than the usually quoted evaporation mass in the case of DM-nucleon interactions. Finally, we compare the neutrino rates at production resulting from capture by electrons and nuclei. This is relevant to evaluate the importance of electron capture in leptophilic scenarios, which will be studied elsewhere~\cite{Garani:2017}.

This paper is organized as follows. In Section~\ref{sec:cs} we describe different types of interactions we consider. In Section~\ref{sec:capture} we review the calculation of the capture rate and compare the results of capture by solar electrons and nuclei. In Section~\ref{sec:distribution} we describe the velocity and radial distribution of DM particles in the Sun once equilibrium is attained and show the resulting temperature in the optically thin regime for the different types of interactions and targets (electrons and nuclei). With this at hand, we write down the expression for the annihilation rate. In Section~\ref{sec:evaporation} we review the calculation of the evaporation rate and illustrate our results. In Section~\ref{sec:evapmass} we compute the minimum testable mass below which evaporation is very effective for the different cases under study. In Section~\ref{sec:rates} we compare the neutrino rates at production obtained for capture by electrons and by nuclei for the different cross sections we consider. Finally, in Section~\ref{sec:summary} we summarize our findings and draw our conclusions. In three appendices we describe the calculation of the differential scattering rates (Appendix~\ref{app:scatteringrates}), the calculation of the temperature in the optically thin regime (Appendix~\ref{app:temperature}) and the calculation of some quantities related to the propagation of DM particles in the Sun, mainly relevant in the conduction limit or optically thick regime (Appendix~\ref{app:prop}).

\section{Scattering cross sections}
\label{sec:cs}

The scattering rates that govern the capture and evaporation rates of DM particles in the Sun scale with the scattering cross section in the Knudsen limit (optically thin regime). For the case of interactions off free electrons, the single-particle total (constant) cross section, which appears in the scattering rates, is simply given by $\sigma_e = \sigma_{e,0}$. However, in the case of interactions off nuclei $i$, depending on the type of interactions, either spin-dependent (SD) or spin-independent (SI), the total (constant) DM-nucleus cross sections, at zero momentum transfer, are given, in terms of the DM-proton and DM-neutron cross sections, by
\begin{eqnarray}
\label{eq:csSD}
\sigma_{i,0}^{\rm SD} & = & \left(\frac{\tilde \mu_{A_i}}{\tilde \mu_p}\right)^2 \, \frac{4 \, (J_i +1)}{3 \, J_i} \, \left| \langle S_{p, i}\rangle + {\rm sign}(a_{\rm p} a_{\rm n}) \left(\frac{\tilde \mu_p}{\tilde \mu_n}\right) \,  \sqrt{\frac{\sigma_{n,0}^{\rm SD}}{\sigma_{p,0}^{\rm SD}}} \, \langle S_{n,i} \rangle \right|^2 \, \sigma_{p,0}^{\rm SD} ~, \\  
\label{eq:csSI}
\sigma_{i,0}^{\rm SI} & = & \left(\frac{\tilde \mu_{A_i}}{\tilde \mu_p}\right)^2 \, \left| Z_i + (A_i-Z_i) \, {\rm sign}(f_p f_n) \, \left(\frac{\tilde \mu_p}{\tilde \mu_n}\right) \, \sqrt{\frac{\sigma_{n,0}^{\rm SI}}{\sigma_{p,0}^{\rm SI}}} \right|^2 \, \sigma_{p,0}^{\rm SI} ~,
\end{eqnarray}
where $\tilde \mu_{A_i}$ ($\tilde \mu_{p/n}$) is the reduced mass of the DM-nucleus $i$ (DM-proton/neutron) system, $\sigma_{p,0}^{\rm SD}$ ($\sigma_{n,0}^{\rm SD}$) and $\sigma_{p,0}^{\rm SI}$ ($\sigma_{n,0}^{\rm SI}$) are the SD and SI elastic scattering DM cross section off protons (neutrons), respectively, $Z_i$, $A_i$ and $J_i$ are the atomic number, the mass number and the total angular momentum of the nucleus $i$, and $\langle S_{p,i}\rangle$ and $\langle S_{n,i}\rangle$ are the expectation values of the spins of protons and neutrons averaged over all nucleons, which we take\footnote{For $^{14}$N, we take $\langle S_{p,^{14}\textrm{N}} \rangle = -0.130$ and $\langle S_{n,^{14}\textrm{N}}\rangle = -0.106$, which we obtain by considering the proton and neutron as if they were the only unpaired nucleon within the odd-group model~\cite{Engel:1989ix}.} from Refs.~\cite{Ellis:1987sh, Pacheco:1989jz, Engel:1989ix, Engel:1992bf, Divari:2000dc} (see Ref.~\cite{Bednyakov:2004xq} for a review). The quantities $a_p$ ($f_p$) and $a_n$ ($f_n$) are the axial (scalar) four-fermion DM-nucleon couplings.  As usually done, we assume $\tilde \mu_n^2 \, \sigma_{p,0}^{\rm SD} = \tilde \mu_p^2 \, \sigma_{n,0}^{\rm SD}$, $\tilde \mu_n^2 \, \sigma_{p,0}^{\rm SI} = \tilde \mu_{p}^2 \, \sigma_{n,0}^{\rm SI}$ and the same sign for the couplings, so Eqs.~(\ref{eq:csSD})  and~(\ref{eq:csSI}) get simplified as 
\begin{eqnarray}
\label{eq:crosssectionsSD}
\sigma_{i,0}^{\rm SD} & = & \left(\frac{\tilde \mu_{A_i}}{\tilde \mu_p}\right)^2 \,
\frac{4 \, (J_i +1)}{3 \, J_i} \, \left| \langle S_{p, i} \rangle +
\langle S_{n,i} \rangle \right|^2 \, \sigma_{p,0}^{\rm SD} ~, \\  
\label{eq:crosssectionsSI}
\sigma_{i,0}^{\rm SI} & = & \left(\frac{\tilde \mu_{A_i}}{\tilde \mu_p}\right)^2 \,
A_i^2 \, \sigma_{p,0}^{\rm SI} ~.
\end{eqnarray}
In the case of SD cross sections, the coupling with protons is the one which is mainly probed because almost all DM interactions are off hydrogen.

However, only in the case of constant cross sections, the scattering rate directly depends on the total cross section. For velocity-dependent and momentum-dependent cross sections, the differential cross section enters the calculation. In this work, in addition to the usual constant (velocity-independent and isotropic) cross section case, we also consider $v_{\rm rel}^2$-dependent (isotropic) and $q^2$-dependent cross sections, where $v_{\rm rel}$ and $q$ are the relative DM-target velocity and the transfer momentum, respectively. The differential cross sections (in the limit of zero transfer momentum\footnote{In the case of interactions with nuclei, when the wavelength corresponding to the transfer momentum $q$ is small compared to the nuclear size, the cross section is suppressed with increasing $q$. This is taken into account by the nuclear form factor, which we discuss in the next section.}) for the constant, $v_{\rm rel}^2$-dependent and $q^2$-dependent cases can be written as
\begin{eqnarray}
\label{eq:dss0}
\frac{\dd \sigma_{i, \rm const} (v_{\rm rel}, \cos \theta_{\rm cm})}{\dd \cos{\theta_{\rm cm}}}  & = & \frac{\sigma_{i,0}}{2} ~, \\
\label{eq:dsvn0}
\frac{\dd \sigma_{i,v_{\rm rel}^2} (v_{\rm rel}, \cos \theta_{\rm cm})}{\dd \cos{\theta_{\rm cm}}} & = & \frac{\sigma_{i,0}}{2} \, \left(\frac{v_{\rm rel}}{v_0}\right)^2 ~, \\
\label{eq:dsqn0}
\frac{\dd \sigma_{i,q^2} (v_{\rm rel}, \cos \theta_{\rm cm})}{\dd \cos{\theta_{\rm cm}}} & = & \frac{\sigma_{i,0}}{2} \, \frac{(1 + \mx/m_i)^2}{2} \, \left(\frac{q}{q_0}\right)^2 ~,
\end{eqnarray}
where $\theta_{\rm cm}$ is the center-of-mass scattering angle, $v_0$ and $q_0$ are a reference relative velocity and transfer momentum, and $\mx$ and $m_i$ are the DM and target $i$ masses, respectively. The mass-dependent term in Eq.~(\ref{eq:dsqn0}) is included so that the total $q^2$-dependent cross section is equal to the $v_{\rm rel}^2$-dependent cross section when $q_0 = \mx \, v_0$ and $\sigma_{i,0}$ is the same in both cases, i.e., $\sigma_{i,v_{\rm rel}^2} = \sigma_{i,q^2} = \sigma_{i,0} \left(v_{\rm rel}/v_0\right)^2$. In this work, we use $v_0 = 220$~km/s. See Appendix~\ref{app:scatteringrates} for further comments, definitions and for a description of how the differential cross sections enter the calculation of the differential scattering rates.

\section{Capture of dark matter by the Sun}
\label{sec:capture}

DM particles from the galactic halo could get eventually captured by the Sun if, after scattering off solar targets (nuclei and electrons), they lose energy so that their resulting velocity is lower than the Sun's escape velocity at a distance $r$ from the center of the Sun, $v_e(r)$. The capture rate of DM particles with mass $\mx$ for weak cross sections, for which the probability of interaction is small, is (to good approximation) given by
\begin{equation}
\label{eq:captureweak}
C_{\odot}^{\rm weak} =\sum_i \int_0^{R_\odot} 4 \pi r^2 \dd r \int_0^\infty \dd u_\chi \, \left(\frac{\rho_\chi}{\mx}\right) \, \frac{f_{v_\odot}(u_\chi)}{u_\chi} \, w(r) \int_0^{v_e(r)} R_i^- (w \to v) \, |F_i(q)|^2 \, \dd v ~, 
\end{equation}
where the sum is over all possible targets. In this work we consider electrons and 29 nuclei as targets and use their density and temperature distributions as determined within the standard solar model~\cite{Asplund:2009fu, Serenelli:2011py} (see Ref.~\cite{Vinyoles:2016djt} for a recent update). The factor $R_i^-(w \to v)$ (and the analogous $R_i^+ (w \to v)$) is the differential scattering rate at which a DM particle with velocity $w$ scatters off a target with mass $m_i$ to a final velocity $v<w$ ($v>w$). They are explicitly given in Appendix~\ref{app:scatteringrates} for constant, $v_{\rm rel}^2$-dependent and $q^2$-dependent cross sections.

The nuclear form factor for nucleus $i$ is $|F_i(q)|^2$, which we approximate as the one corresponding to a Gaussian nuclear density distribution with root-mean-square radius $r_i$ (i.e., equal to that of a uniform sphere of radius $\sqrt{5/3} \, r_i$), i.e., 
\begin{equation}
\label{eq:ff}
|F_i(q)|^2 = e^{- q^2 \, r_i^2/3} ~.
\end{equation}
For SI interactions~\cite{Eder:1968},
\begin{equation}
\label{eq:rSI}
r_i =  \left(0.89 \, A_i^{1/3} + 0.3\right)~\textrm{fm} ~,
\end{equation}
and given that the nuclear density distribution is different from the spin distribution~\cite{Engel:1991wq}, for SD interactions~\cite{Belanger:2008sj},
\begin{equation}
\label{eq:rSD}
r_i = \frac{\sqrt{3}}{2} \, \left(1.7 \, A_i^{1/3} - 0.28 - 0.78 \,  \left(A_i^{1/3} - 3.8 + \sqrt{\left(A_i^{1/3} - 3.8\right)^2 + 0.2}\right)\right)~\textrm{fm} ~. 
\end{equation}
For electrons and hydrogen, $F_e (q^2) = F_H (q^2) = 1$. 

A few comments are in order. Note that a more realistic Woods-Saxon nuclear density distribution (for SI interactions) results in a form factor which is very similar to Eq.~(\ref{eq:ff}) for relatively low $q \, r_i$ values~\cite{Engel:1989ix, Engel:1992bf}. Moreover, Eq.~(\ref{eq:captureweak}) is strictly correct if target nuclei are assumed to be at rest (for electrons and hydrogen, it is always correct as $F_e (q^2) = F_H (q^2)  =1$). In that case: $q^2 = m_i \, \mx \, (w^2 - v^2)$. Otherwise, up-scatterings with a final velocity below the escape velocity must also be considered (a term with the $R_i^+(w \to v)$ factor) and the nuclear form factor cannot be factored out, but has to be included in the calculation of the differential scattering rates $R_i^-(w \to v)$ and $R_i^+(w \to v)$. However, the former correction is negligible and, given the current uncertainties, the fact that we are not using more accurate nuclear response functions~\cite{Anand:2013yka, Catena:2015uha, Gazda:2016mrp} and that in the end factoring out the nuclear form factor represents at most an overall $10 \%$ (much smaller in the constant case) reduction with respect to the results from Eq.~(\ref{eq:captureweak}) for the case of SI interactions only, we do not refine the calculation further and consider the form factor as computed in the zero-temperature limit (but not the differential scattering rates), so that it can be factored out in Eq.~(\ref{eq:captureweak}) and the analytical expressions in Appendix~\ref{app:scatteringrates} can be used. 

The local DM density is given by $\rho_\chi = 0.3$~GeV/cm$^3$, $R_\odot$ is the Sun radius and $f_{v_\odot}(u_\chi)$ is the halo velocity distribution seen by an observer moving at speed $v_\odot$, the speed of the Sun with respect to the DM rest frame,
\begin{equation}
\label{eq:fugen}
f_{v_\odot}(u_\chi)=\frac{1}{2} \, \int_{-1}^{1} f_{\rm gal}(\sqrt{u_\chi^2 + v_\odot^2 + 2 \, u_\chi \, v_\odot \, \cos{\theta_\odot}}) \, \dd\cos{\theta_\odot} ~,
\end{equation}
where $u_\chi$ is the DM velocity at infinity in the Sun's rest frame, $\cos{\theta_\odot}$ is the angle between the DM and the solar system velocities and $f_{\rm gal}(u_{\rm gal})$ is the DM velocity distribution in the galactic rest frame, which is assumed to be a Maxwell-Boltzmann distribution (the so-called standard halo model) and thus, 
\begin{equation}
\label{eq:fu}
f_{v_\odot}(u_\chi) = \sqrt{\frac{3}{2 \, \pi}} \, \frac{u_\chi}{v_\odot \, v_d} \, \left(e^{-\frac{3 \, (u_\chi-v_\odot)^2}{2 \, v_d^2}} - e^{-\frac{3 \, (u_\chi+v_\odot)^2}{2 \, v_d^2}}\right) ~,
\end{equation}
with $w^2(r) = u_\chi^2 + v_e^2(r)$, the square of the DM velocity at a distance $r$ from the center of the Sun.  We take the values $v_\odot = 220$~km/s for the velocity of the Sun with respect to the DM rest frame and thus, $v_d = 270~{\rm km/s} \simeq \sqrt{3/2} \, v_\odot$ for the velocity dispersion. Actually, $f_{\rm gal}(u_{\rm gal})$ does not extend beyond the local galactic escape velocity, $v_{\rm esc, gal} = 533^{+54}_{-41}$~km/s at 90\% confidence level~\cite{Piffl:2013mla}. However, this represents a correction on the capture rate below the percent level~\cite{Choi:2013eda}, much smaller than the very same form of the velocity distribution~\cite{Kundu:2011ek, Danninger:2014xza, Choi:2013eda}. Finally, note that we are assuming the Sun to be in free space, but the presence of the planets (mainly Jupiter) could affect the solar capture rate\footnote{See Refs.~\cite{Gould:1987ww, Gould:1990ad, Damour:1998rh, Damour:1998vg, Gould:1999je, Lundberg:2004dn, Peter:2009mi, Peter:2009mm} for discussions about the effects of the Sun and planets on the DM capture rate by the Earth.}, mainly for heavier DM particles for which the low-velocity tail is more important~\cite{Peter:2009mk}.  Nevertheless, it has been recently shown that planetary diffusion of DM particles in and out of the solar loss cone (orbits crossing the Sun) would result in a complete cancellation of the effect, so the free-space approximation is very accurate, as long as gravitational equilibrium has been reached (in the case of constant scattering cross sections off nucleons, for $\mx=100$~GeV, this occurs for $\sigma^{\rm SD} \gtrsim  10^{-44}~\textrm{cm}^2$ for SD interactions and for $\sigma^{\rm SI} \gtrsim 10^{-46}~\textrm{cm}^2$ for SI interactions)~\cite{Sivertsson:2012qj}. 

\begin{figure}[t]
	\begin{center}
		\includegraphics[width=0.45\linewidth]{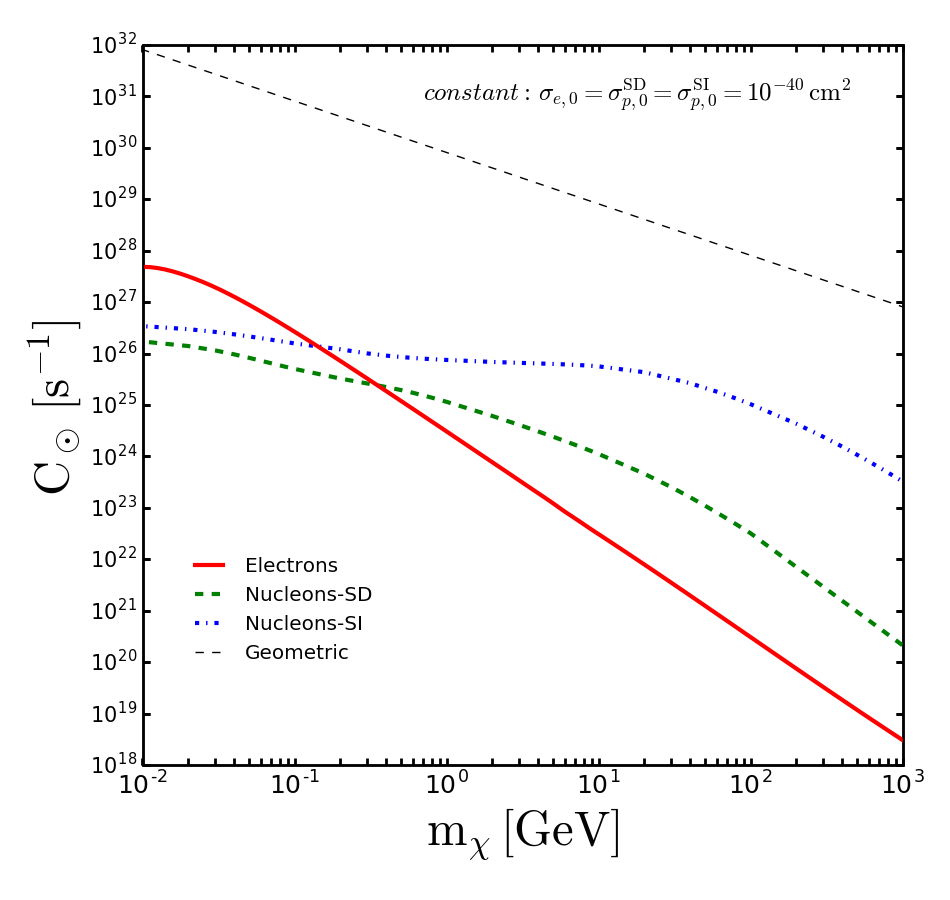}		
		\includegraphics[width=0.45\linewidth]{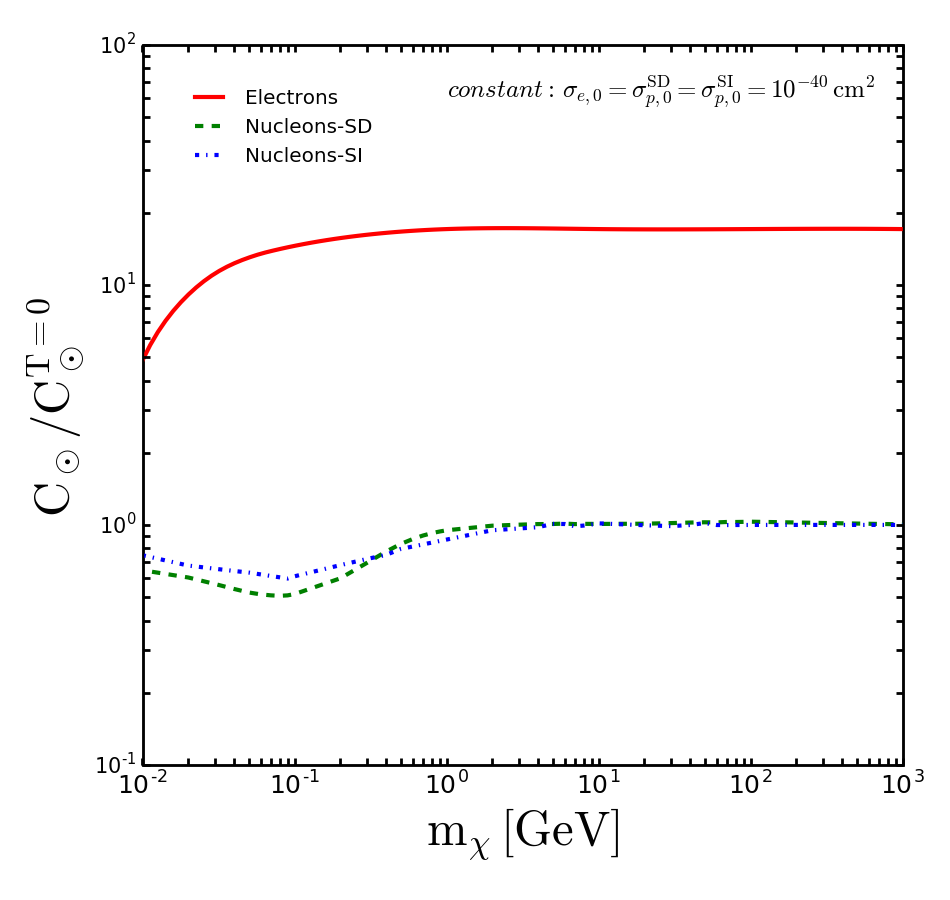} \\
		\includegraphics[width=0.45\linewidth]{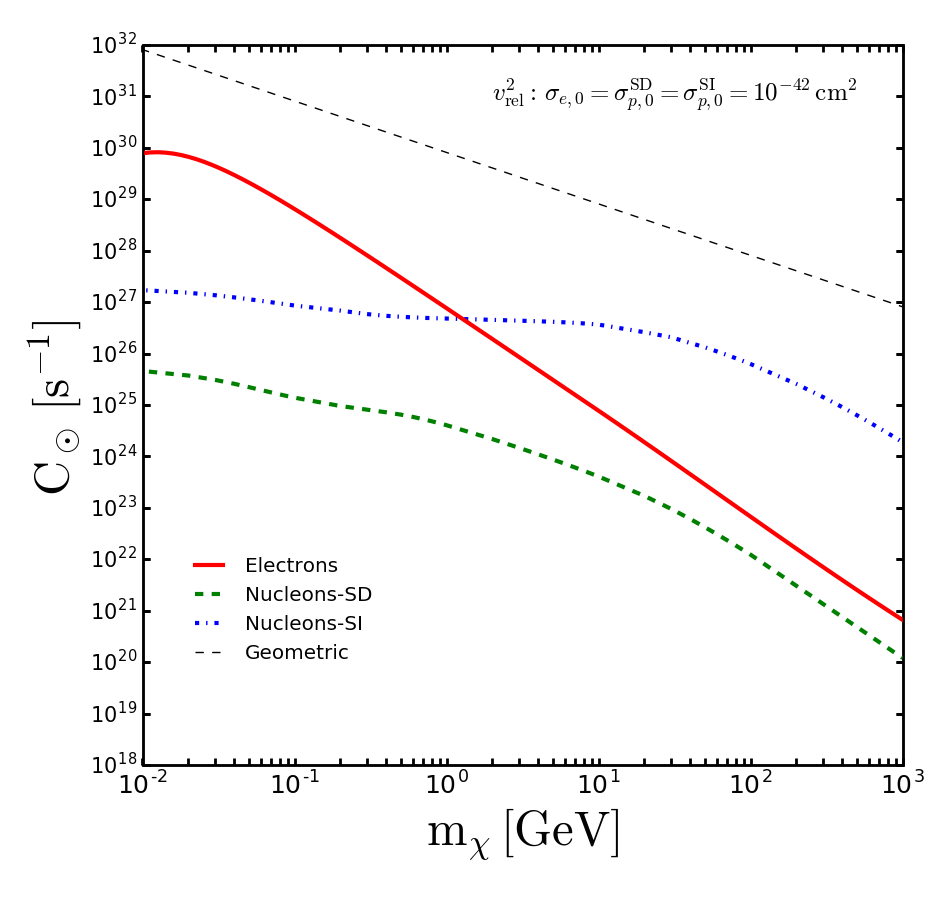}
		\includegraphics[width=0.45\linewidth]{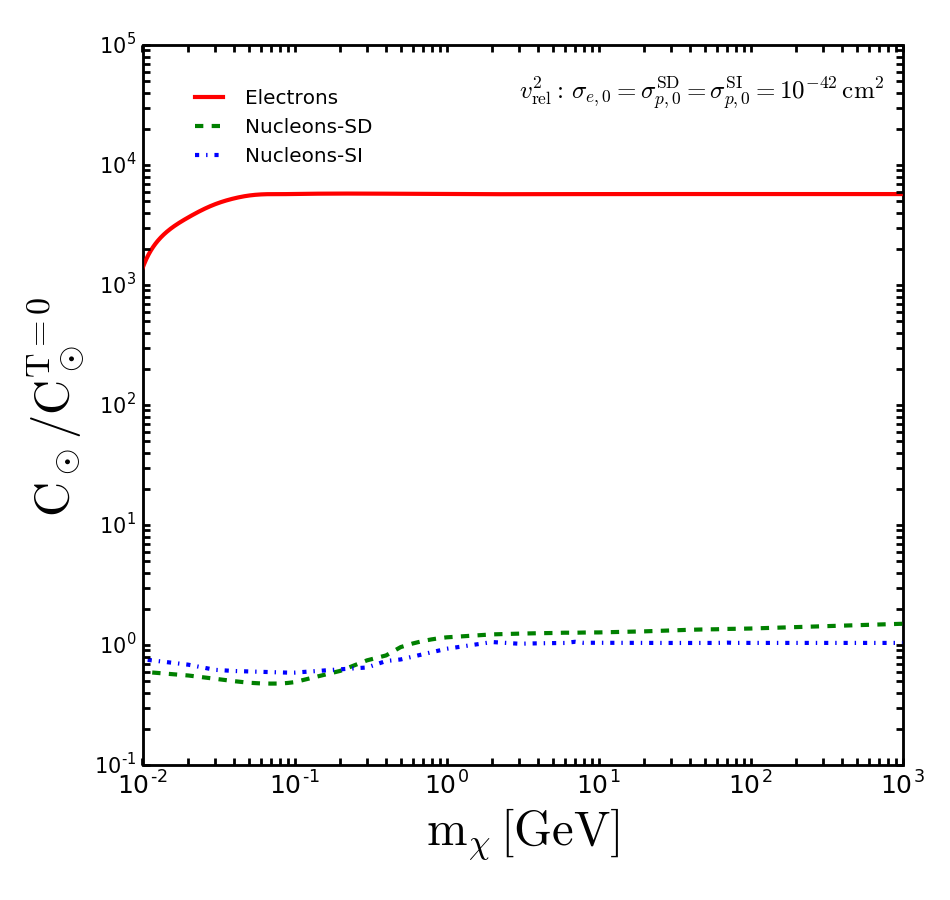} \\
		\includegraphics[width=0.45\linewidth]{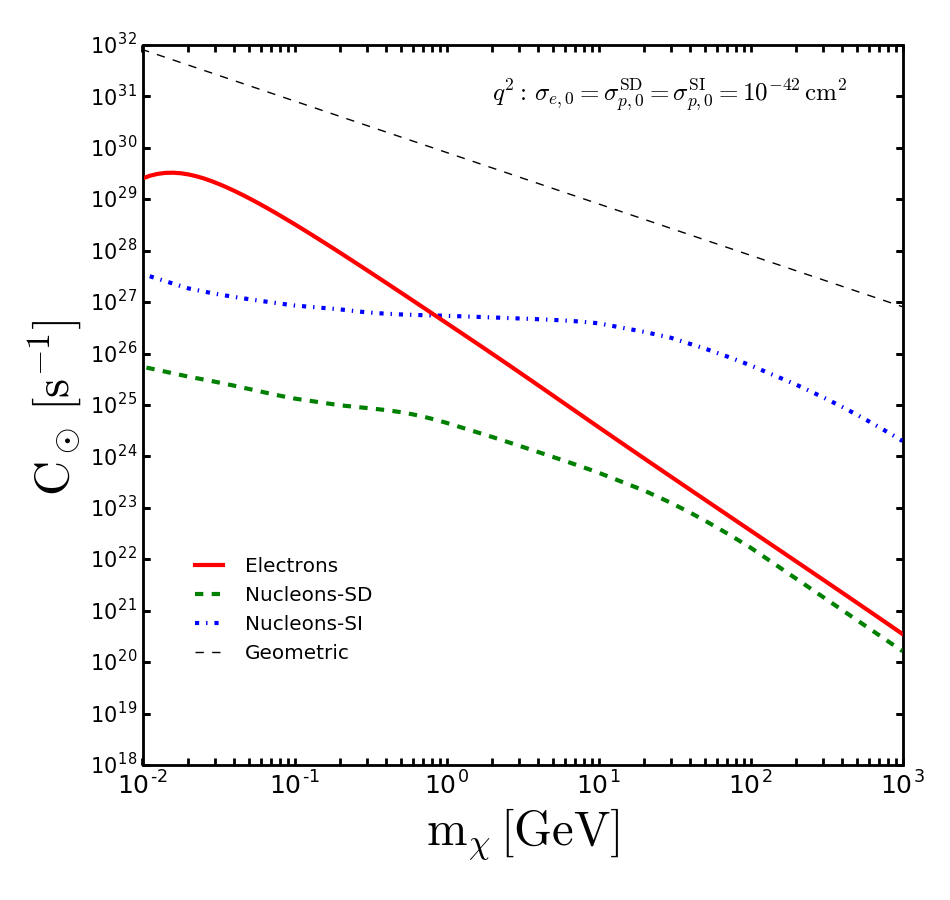}
		\includegraphics[width=0.45\linewidth]{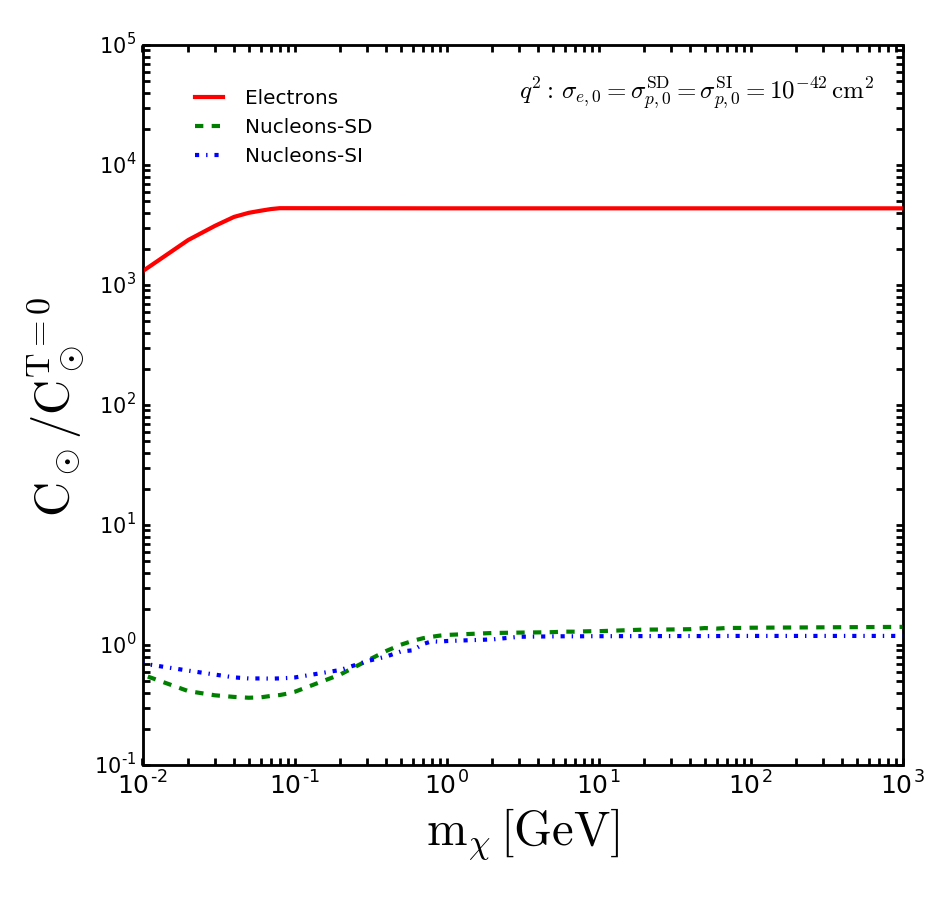}				
	\end{center}
	\caption{\textbf{\textit{Capture rates as a function of the DM mass}}, for DM-electron interactions (solid red curves), DM-nucleon SD interactions (dashed green curves) and DM-nucleon SI interactions (dot-dashed blue curves). {\it Left panels}: capture rates for the three types of interactions. The geometric capture rate is also shown (dashed black curves.) {\it Right panels}: ratio of capture rates with respect to the limit of targets at rest ($T_\odot(r)=0$). {\it Top panels}: constant (velocity-independent and isotropic) scattering cross section with $\sigma_{i,0} = 10^{-40}~\textrm{cm}^2$. {\it Middle panels}: $v_{\rm rel}^2$-dependent scattering cross section with $\sigma_{i,0} = 10^{-42}~\textrm{cm}^2$. {\it Bottom panels}: $q^2$-dependent scattering cross section with $\sigma_{i,0} = 10^{-42}~\textrm{cm}^2$. }
	\label{fig:capture}
\end{figure}

On the other hand, Eq.~(\ref{eq:captureweak}) is only valid for weak scattering cross sections, such that the probability of interaction is very small: the capture rate cannot grow indefinitely with the cross section. The saturation value for the capture rate is set by the geometrical cross section of the Sun (when the probability of interaction and capture is equal to one)~\cite{Bottino:2002pd, Bernal:2012qh},
\begin{equation}
\label{eq:capturegeom}
C_\odot^{\rm geom} = \pi R_\odot^2 \, \left(\frac{\rho_\chi}{m_\chi}\right) \, \int_0^{\infty} \dd u_\chi \, f_{v_\odot}(u_\chi) \, \frac{\omega^2(R_\odot)}{u_\chi} = \pi R_\odot^2 \, \left(\frac{\rho_\chi}{m_\chi}\right)
\, \langle v \rangle_0 \,  \left(1 + \frac{3}{2} \, \frac{v_e^2(R_\odot)}{v_d^2}\right) \xi(v_\odot,v_d) ~, 
\end{equation}
where $\langle v \rangle_0 = \sqrt{8/(3\pi)} \, v_d$ is the average speed in the DM rest frame and the factor $\xi(v_\odot,v_d)$ takes into account the suppression due to the motion of the Sun ($\xi(v_\odot = 0,v_d) = 1$),
\begin{equation}
\label{eq:xi}
\xi(v_\odot,v_d) \equiv \frac{v_d^2 \, e^{- \frac{3 \, v_\odot^2}{2 \, v_d^2}} + \sqrt{\frac{\pi}{6}} \, \frac{v_d}{v_\odot} \, \left(v_d^2 + 3 \, v_e^2(R_\odot) + 3 \, v_\odot^2\right) \, \textrm{Erf}\left(\sqrt{\frac{3}{2}} \, \frac{v_\odot}{v_d}\right)}{2 \, v_d^2 + 3 \, v_e^2(R_\odot)} ~.
\end{equation}
For the chosen values of $v_\odot$ and $v_d$, $\xi(v_\odot = 220 \, \textrm{km/s},v_d = 270 \, \textrm{km/s}) \simeq 0.81$. Finally, in order to allow for a smooth transition between these two regimes, we estimate the capture rate as~\cite{Bernal:2012qh}
\begin{equation}
\label{eq:capture}
C_\odot = C_\odot^{\rm weak} \, \left(1 - e^{-C_\odot^{\rm geom}/C_\odot^{\rm weak}}\right) ~.
\end{equation}
In the left panels of Fig.~\ref{fig:capture}, we show the capture rates as a function of the DM mass for the case of DM-electron interactions (solid red curves), DM-nucleon SD interactions (dashed green curves) and DM-nucleon SI interactions (dot-dashed blue curves), for constant cross sections with $\sigma_{i,0} = 10^{-40}~\textrm{cm}^2$ (top panels), $v_{\rm rel}^2$-dependent cross sections with $\sigma_{i,0} = 10^{-42}~\textrm{cm}^2$ (middle panels) and for cross sections with $q^2$ dependence with $\sigma_{i,0} = 10^{-42}~\textrm{cm}^2$ (bottom panels). In each panel, we also indicate the geometric limit (dashed black curve), Eq.~(\ref{eq:capturegeom}). We stress again that even for leptophilic DM models, in general, interactions with nucleons are possible via loop processes, so the capture rates by nuclei are relevant and need to be considered. 

In the case of constant cross sections (top-left panel), for high DM masses, capture by nuclei is several orders of magnitude (up to two for SD and four for SI) larger than capture by electrons. The differences decrease for lower masses and capture by electrons is comparable or larger for $\mx \lesssim 1$~GeV, which can be relevant if the DM velocity distribution has a cutoff at $v_c(r) < v_e(r)$ (see below). The results for $v_{\rm rel}^2$-dependent and $q^2$-dependent cross sections are similar to each other (for the normalizations used in this work). Unlike for constant cross sections, in these cases, at high masses the capture rate by electrons is a factor of a few larger than capture by nuclei via SD interactions and the differences with respect to the SI capture rate decrease, being of three orders of magnitude. These results can be understood from the even more important impact of thermal effects for these cross sections as compared to the constant case and can have important consequences in some models~\cite{Garani:2017}.  Moreover, the SD capture rate is much smaller than the SI case (up to four orders of magnitude). Overall, the capture rate via $v_{\rm rel}^2$-dependent and $q^2$-dependent cross sections is a factor of about four, three and two orders of magnitude larger than the case with constant cross sections (assuming the same $\sigma_{i, 0}$ for all cases) for capture by electrons, SI and SD interactions off nucleons, respectively.

In the right panels of Fig.~\ref{fig:capture}, we illustrate the impact of thermal effects on the capture rates. These effects are driven by two competing factors: the ratio of the solar temperature to the DM escape energy, $2 \, T_\odot/(\mx \, v_e^2(r)) = u_i^2(r)/(\mu_i \, v_e^2(r))$, where $u_i(r)$ is the most probable speed of the targets at position $r$, and the ratio of the targets thermal speed to the escape velocity. Whereas a larger average kinetic energy of the DM particles suppresses capture, thermal effects enlarge the range of velocities contributing to it. For the same cases of the left panels, we show the ratio of the capture rates obtained using the thermal distribution of the target particles with respect to the capture rates obtained in the limit $T_\odot (r) = 0$, i.e., when the targets are at rest.

In the case of velocity-independent and isotropic cross sections, as discussed in Ref.~\cite{Kopp:2009et} and as can be seen in the top-right panel, thermal effects represent an order of magnitude correction if the target particles are electrons. However, the correction in the case of interactions with nucleons is very small for $\mx \gtrsim 1$~GeV. These differences can be explained by the larger thermal speed of electrons as compared to that for nuclei $i$ by a factor $\sqrt{m_i/m_e}$. In the case of $v_{\rm rel}^2$-dependent and $q^2$-dependent cross sections, thermal effects on the capture rates by electrons are very important and represent an increase of three orders of magnitude in the range of masses we show. This can be understood from the extra $u_i^2(r)/v_0^2 \sim {\rm few \, GeV}/m_i$ factors in the differential scattering rates $R_{i,v_{\rm rel}^2}^-(w \to v)$ and $R_{i,q^2}^-(w \to v)$ (see Appendix~\ref{app:scatteringrates}). For these cross sections, even in the case of capture by nucleons, thermal effects cannot always be neglected. For $\mx \gtrsim 1$~GeV, for DM-nucleon interactions the increase in the capture rate is of a few tens of percent for both, $v_{\rm rel}^2$-dependent and $q^2$-dependent cross sections, although for SI interactions the correction is negligible for $v_{\rm rel}^2$-dependent cross sections. On the other hand, for $\mx \lesssim 1$~GeV, for all cases, thermal effects {\it suppress} the capture rates contrary to the results at higher masses, given that $2 \, T_\odot/(\mx v_e^2(r)) \sim 0.1~{\rm GeV}/\mx$ and $u_i^2(r)/v_e^2(r) \sim 0.1~{\rm GeV}/m_i$ . This explains the dip in the ratios for the case of DM-nucleon interactions at $\mx \sim 0.1$~GeV.

Finally, as mentioned above, we have checked that the correct calculation of the capture rate by nucleons (mainly for SI interactions), i.e., including the form factor in the $R_i^- (w \to v)$ factors, only represents a decrease of $\lesssim 10\%$ with respect to the results shown here.

\section{Dark matter distribution and annihilation rate in the Sun}
\label{sec:distribution}

After DM particles are trapped inside the Sun, successive scatterings with the target material (nuclei and electrons), which is in local thermodynamic equilibrium (LTE), would thermalize them at a temperature $\tx(r)$. Therefore,  the velocity distributions of target and DM particles can be assumed to have a Maxwell-Boltzmann form,
\begin{eqnarray}
\label{eq:fv}
f_i(\boldsymbol{u},r) & = & \frac{1}{\sqrt{\pi^3}} \, \left(\frac{m_i}{2 \, T_\odot(r)}\right)^{3/2} \, e^{- \frac{m_i \,  u^2}{2 \, T_\odot(r)}} ~, \\
\label{eq:DMdistiso}
f_{\chi}(\boldsymbol{w},r) & = & \frac{e^{- w^2/\vx^2(r)} \, \, \Theta(v_c(r) - w)}{\sqrt{\pi^3} \, \vx^3(r) \, \left(\text{Erf}\left(\frac{v_c(r)}{\vx(r)}\right) - \frac{2}{\sqrt{\pi}} \, \frac{v_c(r)}{\vx(r)} \, e^{- v_c^2(r)/\vx^2(r)}\right)} ~,
\end{eqnarray}
where $T_\odot(r)$ and $\vx(r) \equiv \sqrt{2 \, \tx(r)/\mx}$ are the solar temperature and the thermal DM velocity at a distance $r$ from the center of the Sun, respectively. Whereas in the case of large scattering cross sections (conduction limit or optically thick regime), DM particles would also be in LTE with the solar medium, i.e., $\tx(r) = T_\odot(r)$, in the case of weak cross sections (Knudsen limit or optically thin regime), the DM distribution could be approximated as being isothermal, i.e., with a single temperature, $\tx$. 

Note that we have included a cutoff in the DM velocity distribution, $v_c(r)$, which in general depends on the position (a valid assumption for circular orbits) and it is usually assumed to be equal to the escape velocity at a distance $r$ from the center of the Sun, $v_c(r) = v_e(r)$, but we also consider another possibility, $v_c(r) = 0.9 \, v_e(r)$. The last choice is motivated by the fact that the bulk of evaporation and annihilation takes place in the solar core and for DM particles only passing through the core such a cutoff is a reasonable approximation to the actual distribution function~\cite{Gould:1987ju}. As apparent from the comparison of the results of this approximation with those of Refs.~\cite{Gould:1987ju, Liang:2016yjf}, the actual non-thermal distribution (obtained by solving the collisional Boltzmann equation numerically) cannot be accurately mimicked by the approximate radial and truncated velocity distributions assumed in this work (and in most works in the literature). This has already been noted long ago, as the distribution function is locally non-isotropic with radial orbits always dominating and the local temperature in the Knudsen limit is not uniform~\cite{Gould:1989ez}. 

As we will assume, in the case of weak cross sections, after DM particles are captured by the Sun, they would thermalize non-locally by multiple interactions, with a single isothermal (iso) distribution. In this limit (Knudsen limit), their radial distribution can be written as~\cite{Spergel:1984re, Faulkner:1985rm, Griest:1986yu}
\begin{equation}
\label{eq:nDMr}
n_{\chi, {\rm iso}} (r, t) = N_{\chi}(t) \, \frac{e^{-\mx \phi(r)/\tx}}{\int_{0}^{R_\odot}  e^{-\mx \phi(r)/\tx} \, 4\pi r^2 \, \dd r} ~,
\end{equation}
which corresponds to an isothermal sphere following the law of atmospheres, with a radial dependence set by the gravitational potential $\phi(r) = \int_0^r G M_\odot(r')/{r'}^2 \, \dd r'$, with $G$ the gravitational constant and $M_\odot(r)$ the solar mass at radius $r$, and where $N_{\chi} (t)$ is the total population of DM particles at a given time $t$. 

A relatively simple semi-analytical method to treat the problem in the Knudsen limit was proposed in Ref.~\cite{Spergel:1984re}. By assuming a Maxwell-Boltzmann velocity distribution for the DM and target particles, one can obtain a solution to the isothermal assumption by requiring the DM distribution to satisfy its first energy moment and solving for $\tx$. By imposing that there is no net heat transferred between the two gases, the equation to be solved reads~\cite{Spergel:1984re}
\begin{equation}
\label{eq:nonetflow0text}
\sum_i \int_0^{R_\odot} \epsilon_i(r, \tx, v_c) \, 4 \pi r^2 \, \dd r = 0 ~,
\end{equation}
where
\begin{eqnarray}
\label{eq:epstext}
\epsilon_i(r, \tx, v_c) \equiv  \int \dd^3 \boldsymbol{w} \, n_{\chi,{\rm iso}} (r,t_\odot) \, f_{\chi, {\rm iso}}(\boldsymbol{w},r) \int \dd^3 \boldsymbol{u} \, n_i(r) \, f_i(\boldsymbol{u},r)  \, \sigma_{i,0} \, |\boldsymbol{w} - \boldsymbol{u}| \, \langle \Delta E_i \rangle ~,
\end{eqnarray}
is the energy transfer per unit volume and time, with $ \langle \Delta E_i \rangle$ being the energy transfer per collision averaged over the scattering angle and $n_{\chi,{\rm iso}}(r, t_\odot)$ (see Eq.~(\ref{eq:nDMr})) and $n_i(r)$ being the radial distributions of DM particles and targets $i$, respectively. As we mentioned above, this approximation relies on the assumption of a uniform and locally isotropic Maxwell-Boltzmann distribution for the DM particles, conditions which do not hold in a realistic situation~\cite{Gould:1987ju, Gould:1989ez, Liang:2016yjf}. Indeed, the above approximation overestimates the efficiency of energy transfer by a factor of a few, which depends on the DM and target mass ratio~\cite{Gilliland:1986, Nauenberg:1986em, Gould:1989ez}. Baring in mind the approximated nature of this approach, which is the usual one followed in the literature, we also compute the DM distribution function in the Knudsen limit in this way. However, we implement two semi-analytical corrections. First, we perform the calculation with a cutoff in the DM velocity distribution, in order to be consistent with the inputs used for the computation of the annihilation and evaporation rates. Second, we also include the energy flow in the form of evaporated DM particles that escape the Sun, which is relevant for DM masses of a few GeV and below, so that the final equation we solve is
\begin{equation}
\label{eq:newTchi}
\sum_i \int_0^{R_\odot} \epsilon_i(r, \tx, v_c) \, 4 \pi r^2 \, \dd r = \sum_i \int_0^{R_\odot} \epsilon_{{\rm evap}, i}(r, \tx, v_c) \, 4 \pi r^2 \, \dd r  ~,
\end{equation}
where $\epsilon_{{\rm evap}, i}(r, \tx, v_c)$ is defined in Appendix~\ref{app:temperature}. Indeed, when there is a velocity cutoff, in the case of interactions with electrons, unless this correction is included, wrong solutions are found for $\mx \lesssim 1.1$~GeV and $\mx \lesssim 1.5$~GeV and there are no solutions for $\mx \lesssim 0.4$~GeV and $\mx \lesssim 0.5$~GeV, for $v_c(r) = v_e(r)$ and $v_c(r) = 0.9 \, v_e(r)$, respectively. All the relevant expressions for different types of cross sections (velocity-independent and isotropic, velocity-dependent and isotropic and momentum-dependent) and with a generic cutoff in the DM velocity distribution are provided in Appendix~\ref{app:temperature}.

\begin{figure}[t]
	\begin{center}
		\includegraphics[width=0.49\linewidth]{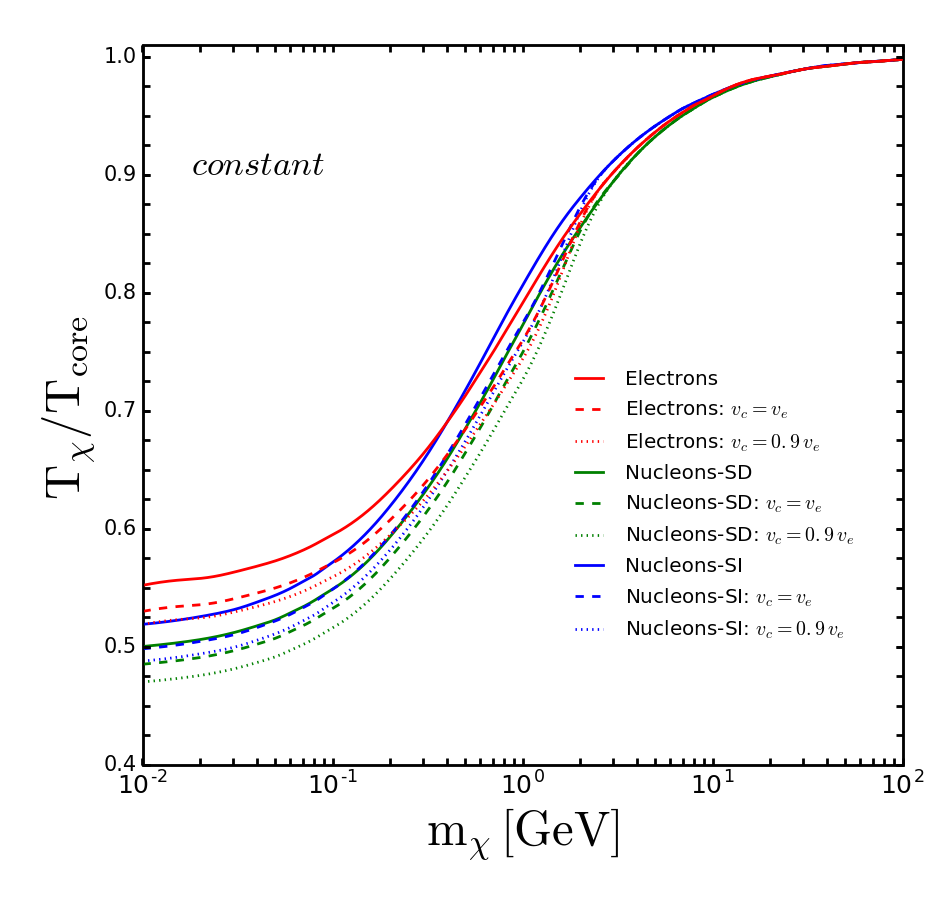} \\		
		\includegraphics[width=0.49\linewidth]{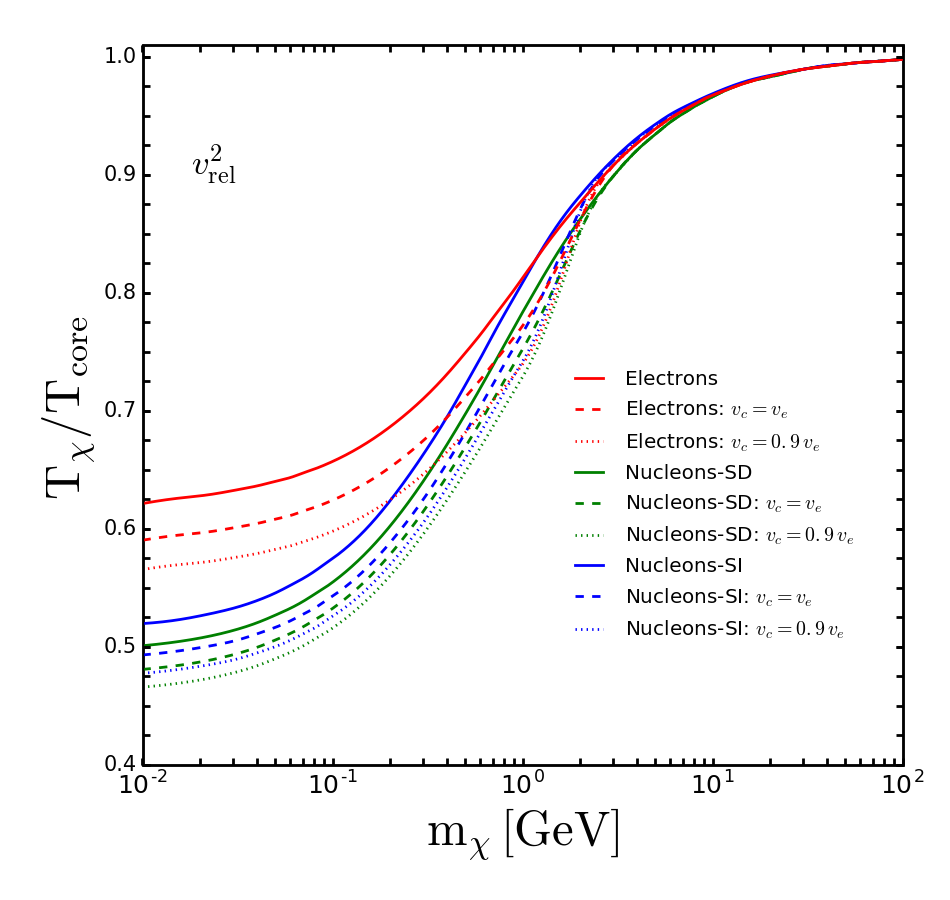}
		\includegraphics[width=0.49\linewidth]{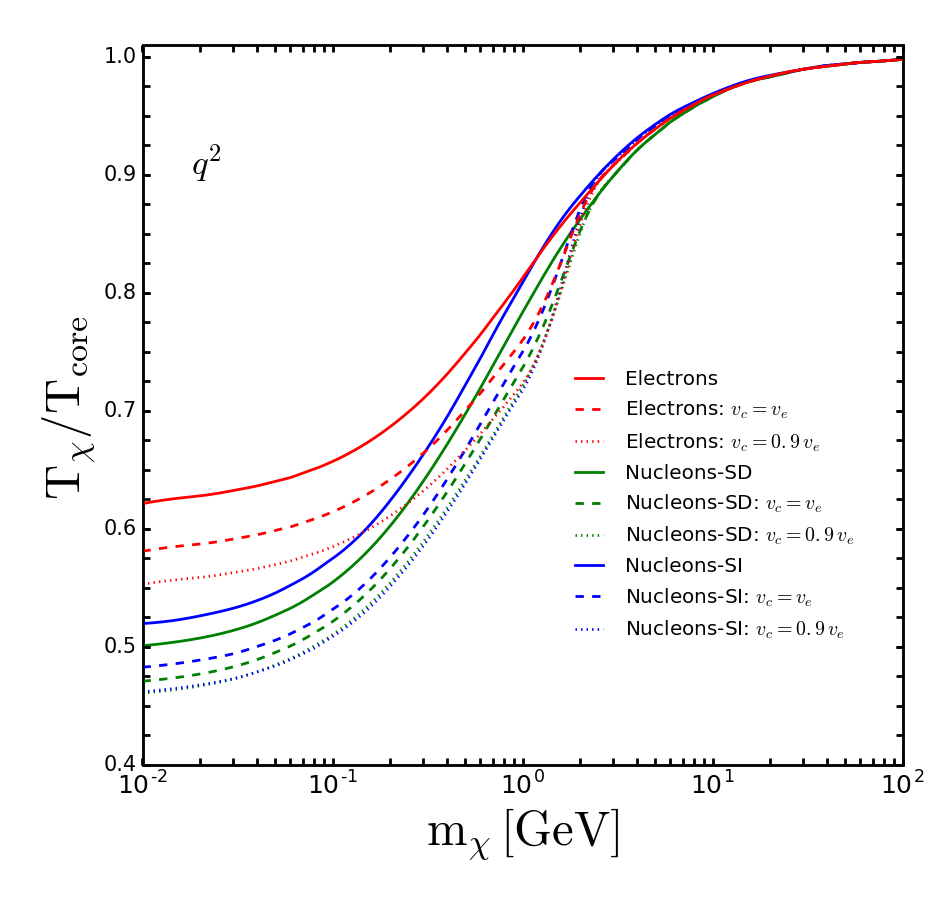}				
	\end{center}
	\caption{\textbf{\textit{DM temperature as a function of the DM mass in the isothermal approximation}}, in units of the solar core temperature, $T_{\rm core} \equiv T_\odot (0)$, for DM scattering off electrons (red curves), off nucleons via SD interactions (green curves) and off nucleons via SI interactions (blue curves), and for three DM velocity distributions: without cutoff (solid curves), with a cutoff at $v_c(r) = v_e (r)$ (dashed curves) and with a cutoff at $v_c (r) = 0.9 \, v_e (r)$ (dotted curves). {\it Top panel}: constant (velocity-independent and isotropic) scattering cross sections. {\it Bottom-left panel}: $v_{\rm rel}^2$-dependent scattering cross sections. {\it Bottom-right panel:} $q^2$-dependent scattering cross sections.}
	\label{fig:temperature}
\end{figure}

In Fig.~\ref{fig:temperature} we show the results for the temperature as a function of the DM mass in the one-zone model or isothermal approximation for electrons (red curves), nucleons with SD (green curves) and SI (blue curves) interactions, for the case of no cutoff, $v_c(r) = \infty$, (solid curves), $v_c(r) = v_e (r)$ (dashed curves) and $v_c (r) = 0.9 \, v_e (r)$ (dotted curves). We show the temperatures for constant (top panel), $v_{\rm rel}^2$-dependent (bottom-left panel) and $q^2$-dependent (bottom-right panel) cross sections. For $\mx \gtrsim 2$~GeV, the temperatures for the three velocity distributions are practically equal, i.e., the cutoff has no effect. This can be understood by the fact that the larger the mass the lower the typical velocities of the DM particles and thus, the high-velocity tail of the distribution is less important. Notice also that all results converge in the large mass limit. For $\mx \lesssim 2$~GeV, the lower the cutoff velocity, the lower the temperature, but the differences are never larger than 10\% for all shown cases. For these low masses, the temperature in the case of interactions off electrons is slightly larger than that obtained when DM interacts with nucleons, being relatively larger for $v_{\rm rel}^2$-dependent and $q^2$-dependent cross sections. Notice also that, in the case of interactions with nucleons, the temperatures for the three cross section dependences are very similar. In the case of thermalization with electrons, constant cross sections result in a bit lower temperatures than the other two cases (up to $\sim 10\%$), the differences getting reduced for low-cutoff velocities. Whereas for the no-cutoff case, for which the correction due to evaporation is negligible, the temperatures for $v_{\rm rel}^2$-dependent and $q^2$-dependent cross sections (as defined in this work) are exactly equal (see Appendix~\ref{app:temperature}), for the case with cutoff, $v_{\rm rel}^2$-dependent cross sections result in slightly larger temperatures. Overall, the differences in the temperatures for the cases under consideration are small for the relevant range of masses and have a small impact on the final neutrino fluxes. 

On the other hand, in the case of large scattering cross sections (conduction limit or optically thick regime), DM particles thermalize locally, i.e., $\tx = T_\odot(r)$, and the DM radial distribution can be approximated as~\cite{Nauenberg:1986em, Gould:1989hm}
\begin{equation}
\label{wq:DMdistLTE}
n_{\chi, {\rm LTE}} (r,t) = n_{\chi, {\rm LTE,0}}(t) \, \left(\frac{T_\odot(r)}{T_\odot(0)}\right)^{3/2} \, {\rm exp}\left(-\int_{0}^{r} \frac{\alpha(r') \frac{\dd T_\odot(r',t)}{\dd r'} + \mx \frac{\dd \phi(r')}{\dd r'}}{T_\odot(r')} \, \dd r'\right) ~,
\end{equation}
where $n_{\chi, {\rm LTE},0}(t)$ is set by the normalization $\int_{0}^{R_\odot} n_{\chi, {\rm LTE}} (r) \, 4\pi r^2 \, \dd r = N_{\chi}(t)$.  For an admixture of targets, a good approximation for the dimensionless thermal diffusivity $\alpha (r)$ is represented by the weighted mean of the single-target solutions~\cite{Gould:1989hm, Gould:1990},
\begin{equation}
\label{eq:alpha}
\alpha(r) = \ell(r) \sum_i \ell_i^{-1}(r) \, \alpha_0 (\mu_i) ~,
\end{equation}
where $\alpha_0(\mu_i)$ is the thermal diffusivity for a single target and it is tabulated as a function of $\mu_i \equiv \mx/m_i$ in Ref.~\cite{Gould:1989hm} for constant cross sections and in Ref.~\cite{Vincent:2013lua} for velocity-dependent and momentum-dependent cross sections. The total mean free path of DM particles in the solar medium is defined as $\ell^{-1}(r) = \sum_i \ell_i^{-1}(r)$, where $\ell_i(r) = (\langle \sigma_i \rangle (r) \, n_i(r))^{-1}$ is the partial mean free path for DM interactions at a distance $r$ from the center of the Sun with a thermal averaged scattering cross section $\langle \sigma_i \rangle (r)$ off targets $i$ with density $n_i(r)$. This thermal average is performed over the DM and target velocity distributions and is given by
\begin{equation}
\label{eq:avgsigma}
\langle \sigma_i \rangle (r) = \int \dd^3 \boldsymbol{w} \, f_{\chi}(\boldsymbol{w},r) \int \dd^3 \boldsymbol{u} \, f_i(\boldsymbol{u},r) \, \sigma_i(\boldsymbol{w},\boldsymbol{u}) ~.
\end{equation}
The expressions for this thermal average for different types of cross sections (constant, velocity-dependent and momentum-dependent) are given in Appendix~\ref{app:mfp}.

The transition from one regime to the other is indicated by the so-called Knudsen number, 
\begin{equation}
\label{eq:Knudsen}
K \equiv \frac{\ell(0)}{r_\chi} ~, \hspace{1cm}
r_\chi = \sqrt{\frac{3 \, T_\odot(0)}{2 \pi G \, \rho_\odot(0) \, \mx}} ~, 
\end{equation}
where $r_\chi$ is the approximate scale height of the DM distribution, with $\rho_\odot(0)$ the density at the solar center.  The Knudsen limit corresponds to $K = \infty$. Note that the definition of the Knudsen number should in principle be a function of the position in the Sun. Nevertheless, given that most of the DM would be concentrated in the center of the Sun, this is sufficient for our purposes and a more accurate definition is beyond the scope of this paper.

Although the actual solution of the problem can only be obtained by solving the collisional Boltzmann equation, however, an approximate solution can be considered by interpolating between the optically thin ($K \gg 1$) and the optically thick ($K \ll 1$) regimes. In order to do so, we follow Refs.~\cite{Bottino:2002pd, Scott:2008ns}, which motivated by the results of Ref.~\cite{Gould:1989hm}, approximated the DM radial and velocity distribution as 
\begin{eqnarray}
\label{eq:DMdist}
n_{\chi} (r, t) \, f_{\chi} (\boldsymbol{w},r) & = & \mathfrak{f}(K) \, n_{\chi, {\rm LTE}} (r, t) \, f_{\chi, {\rm LTE}} (\boldsymbol{w},r) +
\left(1-\mathfrak{f}(K)\right) \, n_{\chi, {\rm iso}} (r, t) \, f_{\chi, {\rm iso}} (\boldsymbol{w},r) ~, ~ ~ ~ ~ ~ ~ ~ \\[1ex]
\label{eq:fKn}
& & \hspace{1.5cm} \mathfrak{f}(K) =  \frac{1}{1+(K/K_0)^2}  ~,
\end{eqnarray}
where $K_0=0.4$ is the value of the Knudsen number for which DM particles transport energy most efficiently~\cite{Gould:1989hm}. This value was obtained by assuming a spherical harmonic oscillator potential and keeping the mean free path as a constant throughout the entire star, which is also the reason why we used the position-independent definition in Eq.~(\ref{eq:Knudsen}). Note that a given $K_0$, which marks the transition from one regime to the other, corresponds to different values of $\sigma_{i,0}$ for different types of cross sections~\cite{Vincent:2015gqa} and targets.

Once the DM distribution is known, we can compute the annihilation rate $A_\odot$, defined as
\begin{equation}
\label{eq:annihilationrate}
A_\odot =  \frac{\int \dd^3 \boldsymbol{w_1} \int \dd^3 \boldsymbol{w_2} \, \sigma_A v_{\chi \chi} \, \int_0^{R_\odot} n_{\chi}(r,t) \, f_\chi(\boldsymbol{w_1},r) \, n_{\chi}(r,t) \, f_\chi(\boldsymbol{w_2},r) \, 4\pi r^2 \, \dd r }{\left(\int_0^{R_\odot} \, n_\chi(r,t) \, 4\pi r^2 \, \dd r \right)^2}  ~,
\end{equation}
where we have used
\begin{equation}
\label{eq:nr}
\int \dd^3 \boldsymbol{w} \, n_{\chi} (r, t) \, f_{\chi} (\boldsymbol{w},r) = n_{\chi} (r, t) = \mathfrak{f}(K) \, n_{\chi, {\rm LTE}} (r, t) + \left(1-\mathfrak{f}(K)\right) \, n_{\chi, {\rm iso}} (r, t)
\end{equation}
in the denominator and where $\sigma_A v_{\chi \chi}$ is the DM annihilation cross section times the relative velocity of the two DM particles, $v_{\chi \chi} = |\boldsymbol{w_2} - \boldsymbol{w_1}|$. In general, $\sigma_A v_{\chi \chi} = a + b \, v_{\chi \chi}^2$, but in this work, our default case is that of an $s$-wave annihilation cross section corresponding to a thermal DM candidate, i.e., $\langle \sigma_A v_{\chi \chi}\rangle = 3 \times 10^{-26} \, \rm{cm}^3/\rm{s}$, where $\langle \, \rangle$ denotes thermal average over the two DM velocity distributions. In such a case, Eq.~(\ref{eq:annihilationrate}) simplifies as
\begin{equation}
\label{eq:annihilationratesim}
A_\odot =  \langle \sigma_A v_{\chi \chi}\rangle \, \frac{\int_0^{R_\odot} n_\chi^2(r,t) \, 4\pi r^2
	\, \dd r }{\left(\int_0^{R_\odot} \, n_\chi(r,t) \, 4\pi r^2 \, \dd r \right)^2}  ~.
\end{equation}
Note that for $p$-wave annihilations, for $v_{\rm rel}^2$-dependent and $q^2$-dependent cross sections equilibrium would be attained for smaller values of $\sigma_{i,0}$ than for the constant case, but we will not discuss this possibility here.

\section{Evaporation rate of dark matter from the Sun}
\label{sec:evaporation}

In general, for sufficiently small DM masses, below a few GeV, interactions with the targets of the solar medium would bring most of the DM particles to velocities above the escape velocity $v_e(r)$, so that they can evaporate from the Sun. The evaporation rate is given by
\begin{equation}
\label{eq:ev+}
E_\odot = \sum_i \int_0^{R_\odot} s(r) \, n_\chi(r,t) \, 4\pi r^2 \, \dd r \, \int_0^{v_c(r)} f_\chi(\boldsymbol{w}, r) \, 4 \pi w^2 \, \dd w \, \int_{v_e(r)}^{\infty} R_i^+ (w \rightarrow v) \dd v ~.
\end{equation}
where the factor $s(r)$ accounts for the suppression of the fraction of DM particles that, even after acquiring a velocity larger than the escape velocity, would actually escape from the Sun due to further interactions in their way out, and can be written as~\cite{Gould:1990} 
\begin{equation}
\label{eq:s}
s(r) = \eta_{\rm ang}(r) \, \eta_{\rm mult}(r) \, e^{-\tau(r)} ~,
\end{equation}
where $\tau(r) = \int_r^{R_\odot} \ell^{-1}(r') \, \dd r'$ is the optical depth at radius $r$. The factors $\eta_{\rm ang} (r)$ and $\eta_{\rm mult}(r)$, which take into account that DM particles travel in non-radial trajectories and that multiple scatterings are possible, are described in Appendix~\ref{app:supp}. Although the result for the factors in $s(r)$ is based on a calculation for a velocity-independent and isotropic cross section~\cite{Gould:1990}, lacking a better estimate, we also use it for the other cases under study. In the optically thin regime, the suppression factor $s(r)$ is nearly one, but we always included it in the calculations.

Note that, to keep it general, we should have considered a term with $R_i^- (w \to v)$ corresponding to down-scatterings to velocities above the escape velocity and hence, the limits for the $R_i^+(w \to v)$ and $R_i^- (w \to v)$ integrals would be $(v_e(r),w)$ and $(w,\infty)$, respectively. Moreover,  a priori, the nuclear form factor for the case of interactions off nuclei must be included too. Whereas in the $R_i^- (w \to v)$ term, the nuclear form factor can be factored out by computing it in the zero-temperature limit, in that limit, the contribution from the $R_i^+(w \to v)$ term is exactly zero. However, for non-zero temperatures, the form factor has to be included in the calculation of the $R_i^+(w \to v)$ term, so an analogous simplification to the one for the $R_i^- (w \to v)$ term cannot be made. Nevertheless, if the DM velocity distribution has a cutoff at $v_c(r) \le v_e(r)$, the $R_i^- (w \to v)$ term is absent and, for the case of the Sun, the nuclear form factors can be approximated to one (at these velocities $q \, r_i \ll 1$). In such cases, which are the ones we consider in this work, the evaporation rate is given by the usual expression, Eq.~(\ref{eq:ev+}).

\begin{figure}[t]
	\begin{center}
		\includegraphics[width=0.45\linewidth]{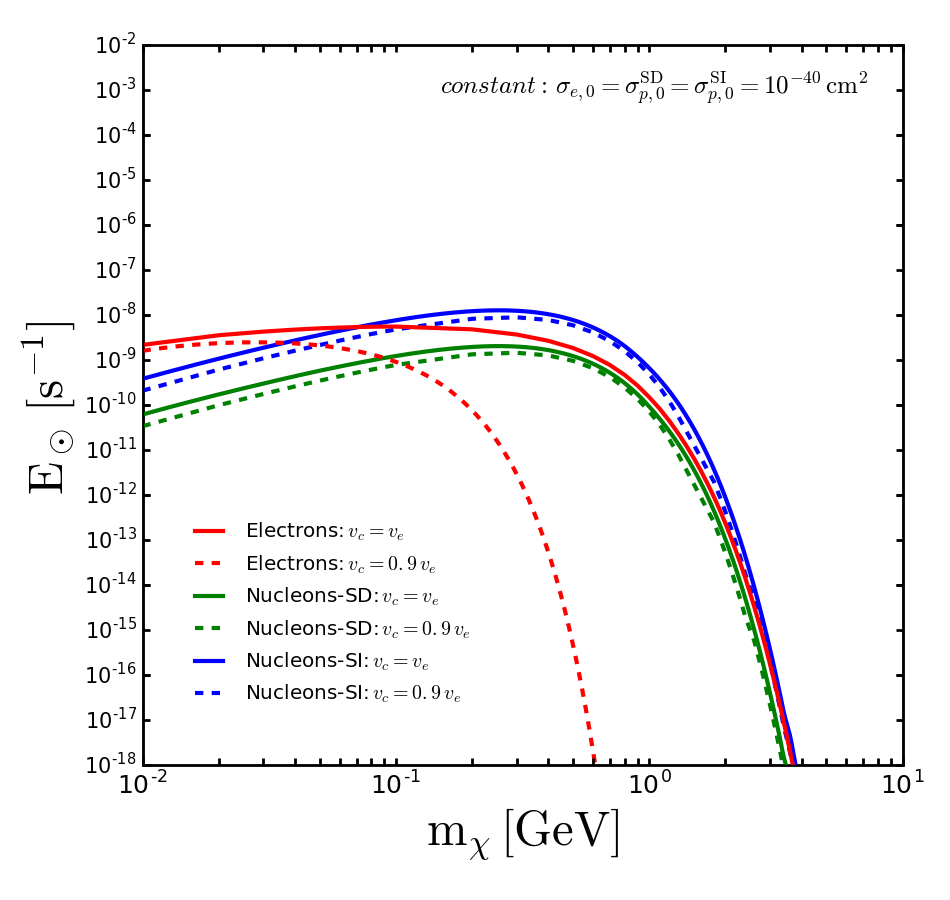}		
		\includegraphics[width=0.45\linewidth]{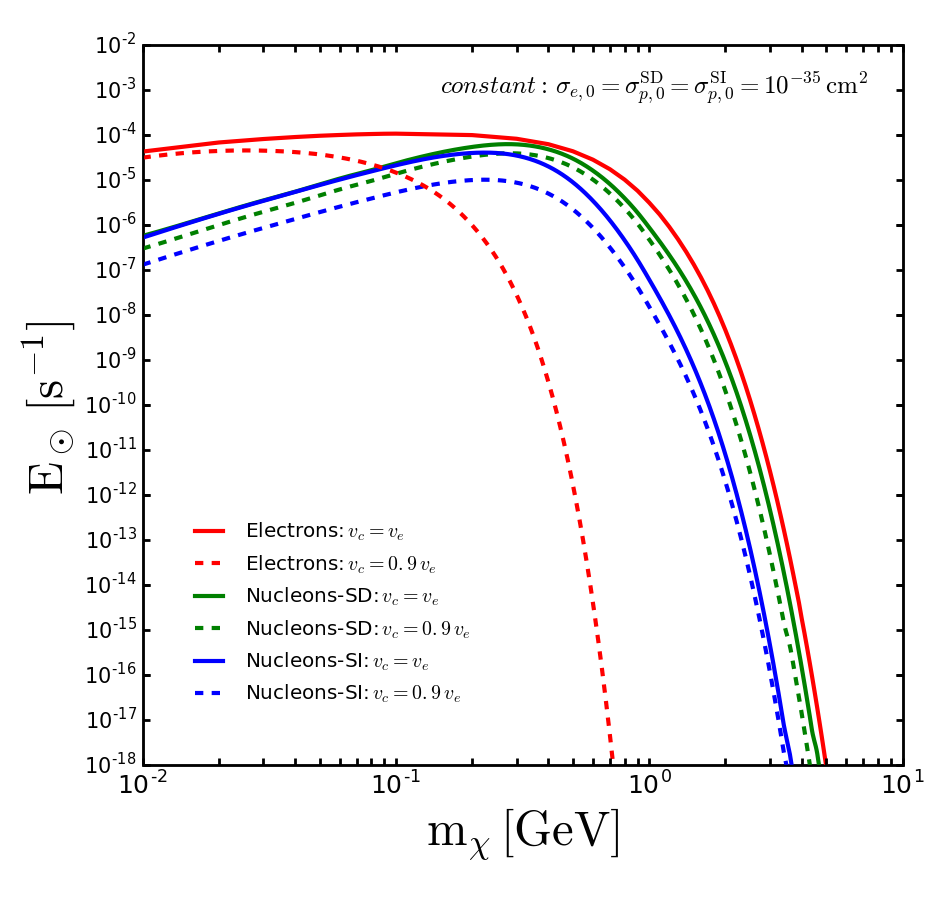} \\	\includegraphics[width=0.45\linewidth]{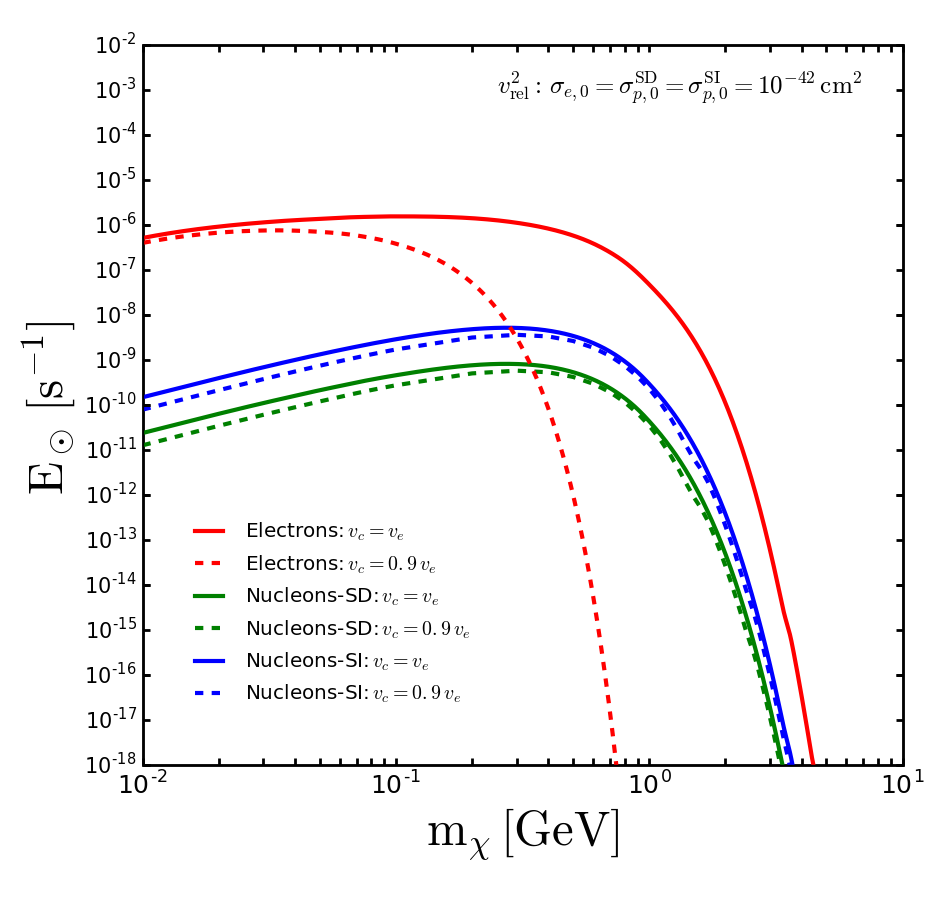}
		\includegraphics[width=0.45\linewidth]{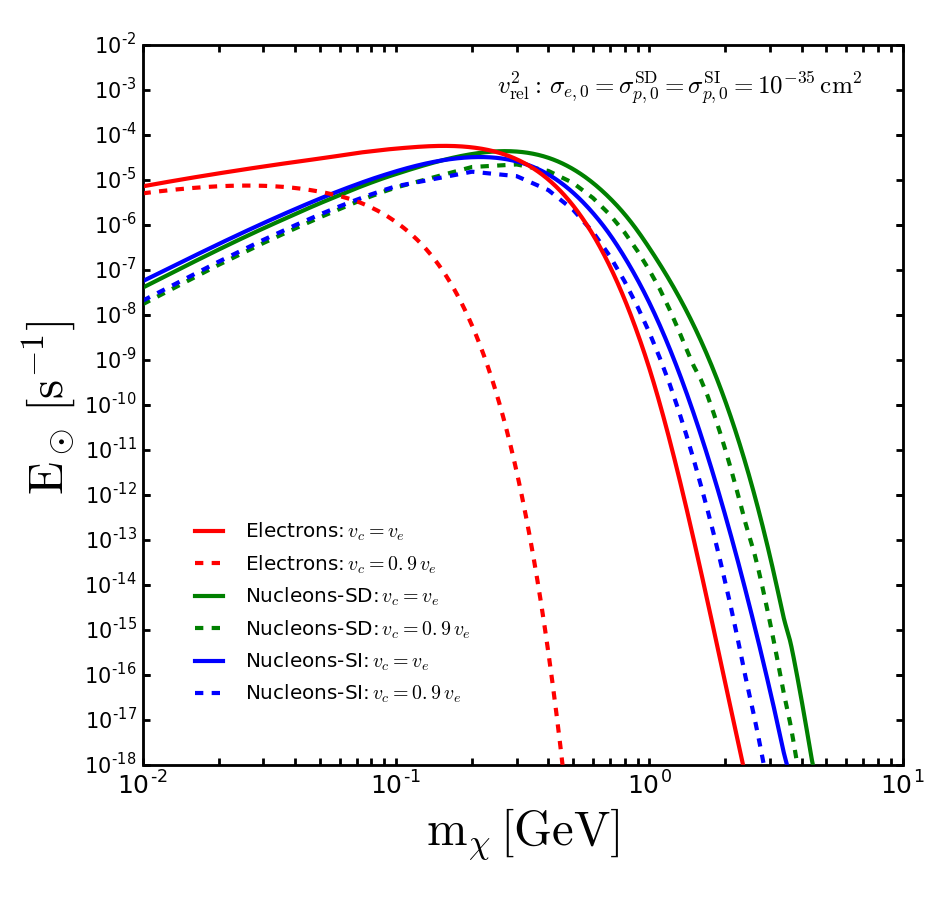} \\		\includegraphics[width=0.45\linewidth]{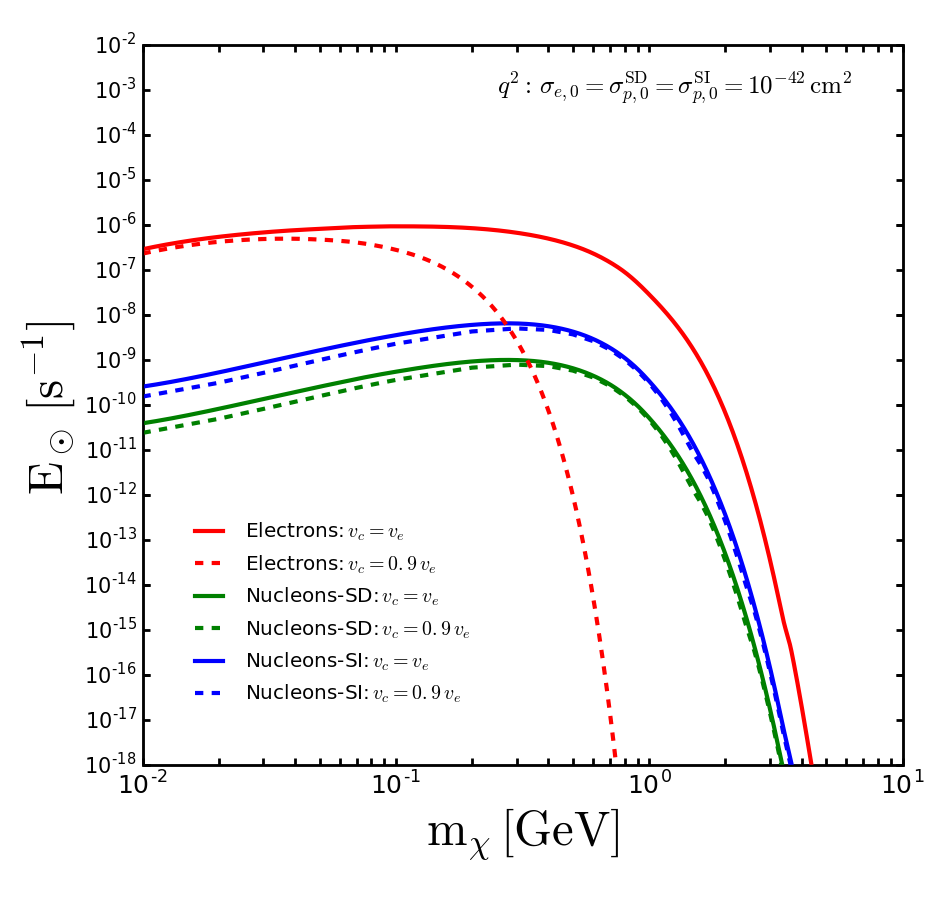}
		\includegraphics[width=0.45\linewidth]{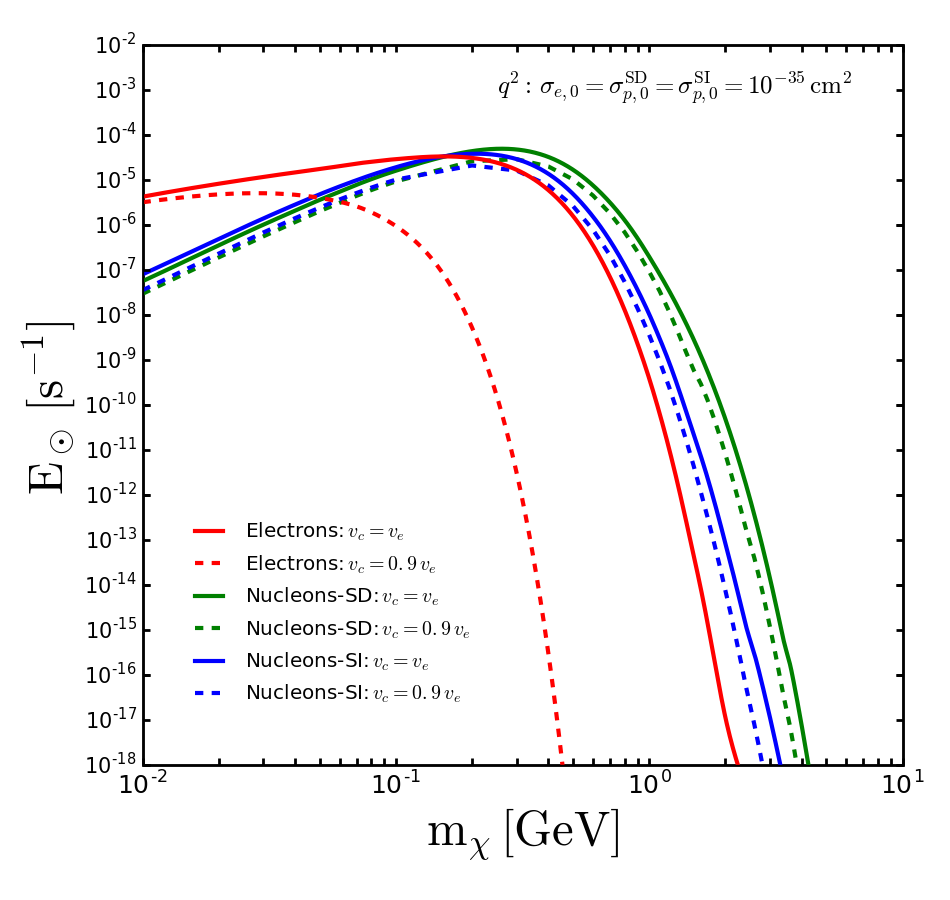}				
	\end{center}
	\caption{\textbf{\textit{Evaporation rates as a function of the DM mass}}, for DM-electron interactions (red curves), DM-nucleon SD interactions (green curves) and DM-nucleon SI interactions (blue curves), with a velocity cutoff at $v_c(r) =v_e(r)$ (solid curves) and at $v_c(r) = 0.9 \, v_e(r)$ (dashed curves). {\it Left panels:} optically thin regime. {\it Right panels:} optically thick regime. {\it Top panels}: constant (velocity-independent and isotropic) scattering cross section for $\sigma_{i,0} = 10^{-40}~\textrm{cm}^2$ and $\sigma_{i,0} = 10^{-35}~\textrm{cm}^2$. {\it Middle panels}: $v_{\rm rel}^2$-dependent scattering cross section for $\sigma_{i,0} = 10^{-42}~\textrm{cm}^2$ and $\sigma_{i,0} = 10^{-35}~\textrm{cm}^2$. {\it Bottom panels}: $q^2$-dependent scattering cross section for $\sigma_{i,0} = 10^{-42}~\textrm{cm}^2$ and $\sigma_{i,0} = 10^{-35}~\textrm{cm}^2$.}
	\label{fig:evaporation}
\end{figure}

In Fig.~\ref{fig:evaporation} we show the evaporation rates for constant (top panels), $v_{\rm rel}^2$-dependent (middle panels) and $q^2$-dependent (bottom panels) cross sections, in the case of DM-electron (red curves), DM-nucleon SD (green curves) and DM-nucleon SI (blue curves) interactions. We consider the same cases depicted in Fig.~\ref{fig:capture}, i.e., cross sections in the optically thin regime (left panels), but also show the results for large cross sections in the optically thick regime (right panels). In all the panels we show the results of a DM velocity distribution with a cutoff at $v_c(r) = v_e (r)$ (solid lines) and $v_c(r) = 0.9 \, v_e (r)$ (dashed lines). The usual exponential fall off at large masses, due to the dependence of the evaporation rate per unit volume on $e^{-\mx v_e^2(r)/2 \tx}$, is clearly visible. Besides this, there are a number of other features worth noticing. 

In the case of interactions with electrons, the effect of the modification of the high-velocity tail of the DM distribution is striking, with a huge impact for masses above ${\cal O}$(0.1)~GeV, for which evaporation is very suppressed in the case of $v_c(r) = 0.9 \, v_e(r)$. On the other hand, in the case of interactions with nucleons, the evaporation rates are only moderately modified when a different cutoff in the DM velocity distribution is considered. This can be understood by the mass scales involved in the problem. Electrons, being light compared to the DM particles, carry little momentum, so that in the case of DM-electron interactions only DM particles with velocities close to the escape velocity are susceptible of gaining enough energy to escape after one interaction. In practice, the differential scattering rate $R_i^+(w \to v)$ for interactions with electrons peaks at DM velocities $w$ close to the escape velocity, whereas for the case of scattering off nucleons it has a broader shape. Therefore, a maximum velocity smaller than the escape velocity significantly suppresses evaporation in the former case. This has important consequences on the available parameter space, as we will discuss below. 

Moreover, whereas in the Knudsen limit the evaporation rate scales linearly with the scattering cross section, for large cross sections, the suppression of evaporation results in a slower-than-linear increase with the cross section. Therefore, although in the optically thin regime the evaporation rate for SI interactions is larger than for SD interactions due to the coherent enhancement of the cross section in the first case, this very same enhancement implies a shorter mean free path and hence, a larger suppression of the evaporation rate for SI cross sections in the optically thick regime. The relative suppression of the rate in the case of interactions with electrons is similar to the case of DM-nucleon SD interactions. We also note that the behavior of the evaporation rates in the case of $v_{\rm rel}^2$-dependent cross sections is very similar to that of $q^2$-dependent cross sections. We find that, in these two cases, the evaporation rates corresponding to DM interactions with electrons are larger than for scatterings off nucleons. In addition, although in general low masses enter the optically thick regime for smaller cross sections (see Eq.~(\ref{eq:Knudsen})), for $v_{\rm rel}^2$-dependent and $q^2$-dependent cross sections this effect is more pronounced (see Eqs.~(\ref{eq:avgsigv2}) and~(\ref{eq:avgsigq2})).

\section{Evaporation mass}
\label{sec:evapmass}

Once all the ingredients are computed, we are interested in knowing what is the minimum DM mass which is testable,\footnote{Note that the high-energy tail of the evaporating DM particles could also be used to probe masses below the evaporation mass~\cite{Kouvaris:2015nsa}.} i.e., which is the minimum DM mass for which DM particles are not evaporated. In order to determine this mass we first consider the evolution of the total number of DM particles in the Sun, which is governed by the following equation: 
\begin{equation}
\label{eq:evolution}
\dot{N_\chi} (t) = C_\odot - A_\odot \, N_\chi^2(t) - E_\odot \, N_\chi (t) ~.
\end{equation}
The solution of this equation, computed at the present time ($t=t_\odot=4.57$~Gyr), is given by~\cite{Gaisser:1986ha, Griest:1986yu}
\begin{equation}
\label{eq:NDM}
N_\chi (t_\odot) = \sqrt{\frac{C_\odot}{A_\odot}} \, \frac{\tanh(\kappa \, t_\odot/\tau_{\rm eq})}{\kappa + \frac{1}{2} \, E_\odot \, \tau_{\rm eq} \, \tanh(\kappa \, t_\odot/\tau_{\rm eq})} ~, 
\end{equation}
where $\tau_{\rm eq} = 1/\sqrt{A_\odot C_\odot}$ is the equilibration time scale in the absence of evaporation and $\kappa = \sqrt{1 + (E_\odot \, \tau_{\rm eq}/2)^2}$. For the usual value assumed for the thermal annihilation cross section, $\langle \sigma_A v_{\chi\chi} \rangle = 3 \times 10^{-26}~{\rm cm}^3/{\rm s}$, and for $\sigma_{e,0} \gtrsim 10^{-42}~{\rm cm}^2$,  $\sigma_{p,0}^{\rm SD} \gtrsim 10^{-43}~{\rm cm}^2$ and $\sigma_{p,0}^{\rm SI} \gtrsim 10^{-44}~{\rm cm}^2$ for constant cross sections (much smaller for the other cases), equilibrium is reached ($\kappa \, t_\odot \gg \tau_{\rm eq}$, $\tanh(\kappa \, t_\odot/\tau_{\rm eq}) \simeq 1$). In the limit when evaporation is important, $\kappa \gg 1$, $N_\chi \simeq C_\odot/E_\odot$, and the number of accumulated DM particles decreases exponentially with decreasing mass (in the optically thin regime). In the limit when evaporation is negligible, $\kappa \simeq 1$, $N_\chi \simeq \sqrt{C_\odot/A_\odot}$, and the number of accumulated DM particles decreases with increasing mass (as $\mx^{-7/4}$ for large masses). Given these considerations, one can define the minimum testable mass or evaporation mass\footnote{For the relevant range of scattering cross sections, other possible, equally good, definitions for the evaporation mass are: $\dd N_\chi / \dd \mx (m_{\rm evap}) = 0$; or $\left|N_\chi (m_{\rm evap}) - \sqrt{\frac{C_\odot (m_{\rm evap})}{A_\odot (m_{\rm evap})}} \, \tanh(t_\odot/\tau_{\rm eq})\right| = \alpha \, N_\chi (m_{\rm evap})$, when the departure from the $E_\odot =0$ limit is of a factor of $\alpha$ (it coincides with Eq.~(\ref{eq:evapmass2}) for $\alpha = \sqrt{11}-1$ and $t_\odot \gg \tau_{\rm eq}$).} as that for which the number of captured DM particles approaches $C_\odot/E_\odot$ at the 10\% level~\cite{Busoni:2013kaa}
\begin{equation}
\label{eq:evapmass}
\left|N_\chi (m_{\rm evap}) - \frac{C_\odot (m_{\rm evap})}{E_\odot (m_{\rm evap})}\right| = 0.1\, N_\chi (m_{\rm evap}) ~, 
\end{equation}
In the limit when equilibrium has been reached, i.e., $\kappa \, t_\odot \gg \tau_{\rm eq}$, it can be written as
\begin{equation}
\label{eq:evapmass2}
E_\odot (m_{\rm evap}) \, \tau_{\rm eq} (m_{\rm evap}) = \frac{1}{\sqrt{0.11}} ~.
\end{equation}
In the optically thin regime, this results in the evaporation mass to increase with the scattering cross section and to decrease with the annihilation cross section. However, when the scattering cross section is large enough and the suppression factor $s(r)$ becomes important, the evaporation mass {\it decreases} with the scattering cross section, so the minimum testable DM mass attains a maximum value around the transition between the two regimes~\cite{Gould:1990, Bernal:2012qh}. Note that for very small scattering cross sections, equilibrium is not reached, the effects of evaporation are negligible and the number of DM particles scales as $N_\chi \simeq C_\odot \, t_\odot$. In this case, there is no evaporation mass, but the total number of DM particles in the Sun would be too small to give rise to any measurable signal.

\begin{figure}[t]
	\begin{center}
		\includegraphics[width=0.49\linewidth]{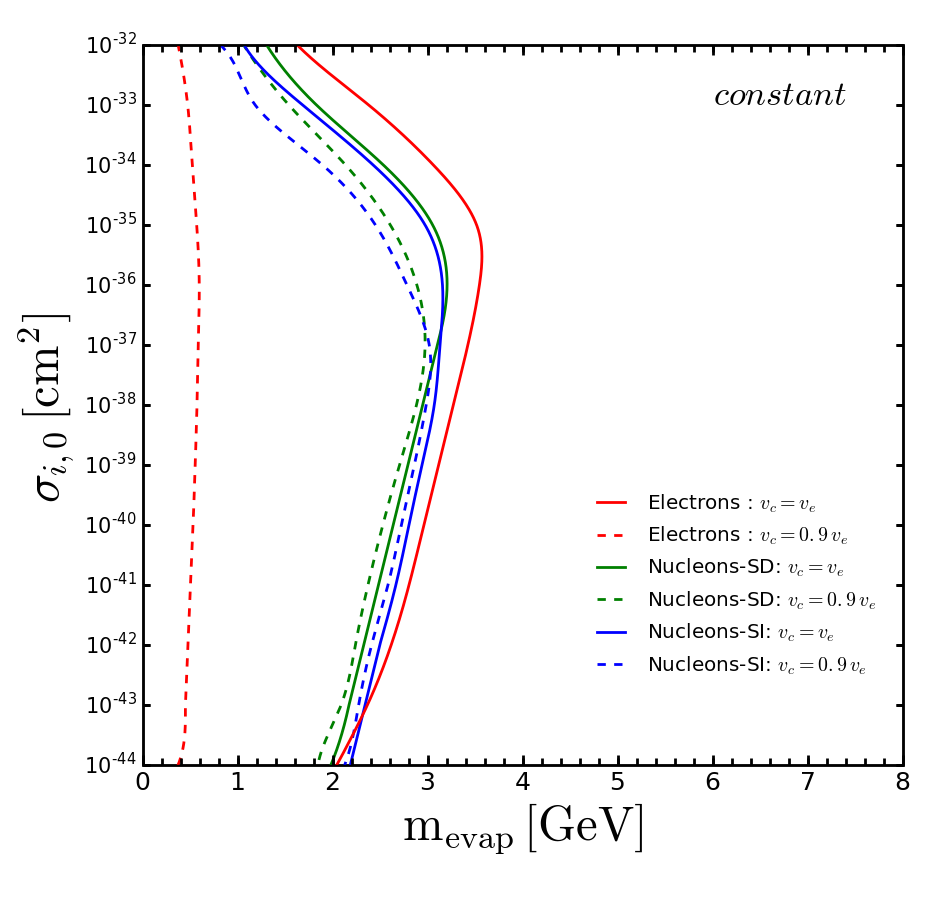} \\		
		\includegraphics[width=0.49\linewidth]{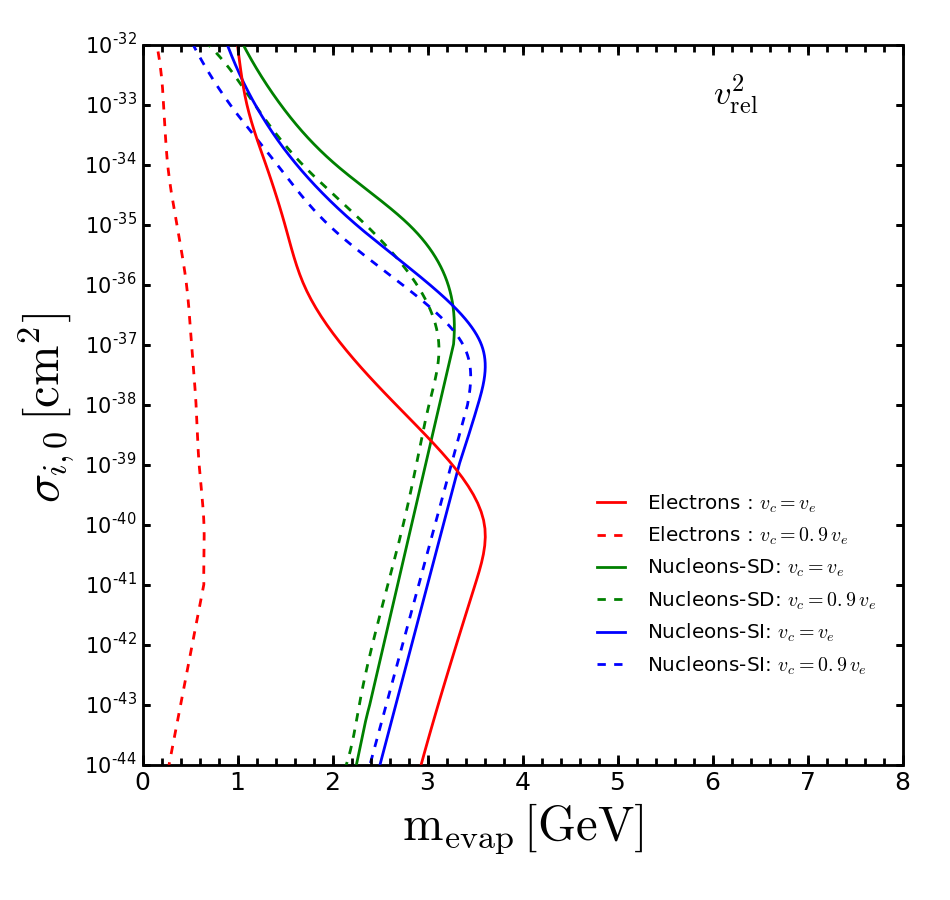}
		\includegraphics[width=0.49\linewidth]{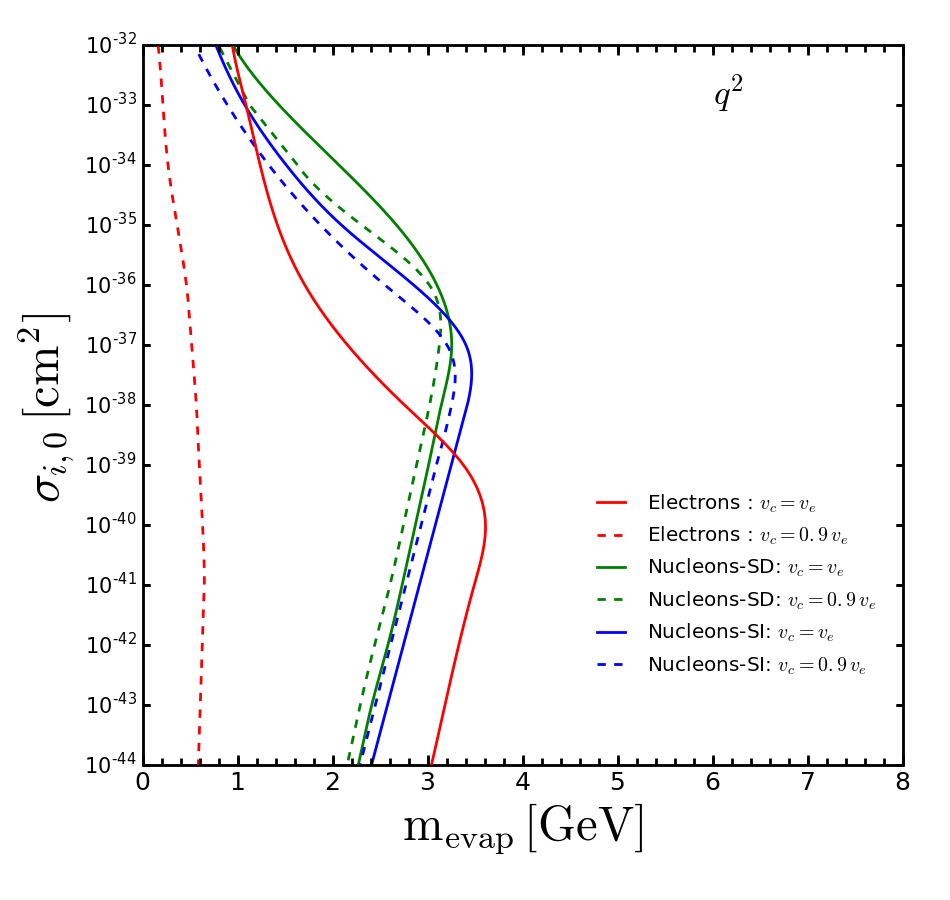}				
	\end{center}
	\caption{\textbf{\textit{Evaporation mass as a function of the scattering cross section}} (in reverse order). Same cases as in Fig.~\ref{fig:evaporation}, i.e., DM scattering off electrons (red curves), off nucleons via SD interactions (green curves) and off nucleons via SI interactions (blue curves) and for two DM velocity distributions: with a cutoff at $v_c(r) = v_e (r)$ (solid curves) and with a cutoff at $v_c (r) = 0.9 \, v_e (r)$ (dashed curves). {\it Top panel}: constant (velocity-independent and isotropic) scattering cross sections. {\it Bottom-left panel}: $v_{\rm rel}^2$-dependent scattering cross sections. {\it Bottom-right panel:} $q^2$-dependent scattering cross sections.}
	\label{fig:evapmass}
\end{figure}

In Fig.~\ref{fig:evapmass} we show (in reverse order) the evaporation mass as a function of the scattering cross section $\sigma_{i, 0}$ for the same cases depicted in Fig.~\ref{fig:evaporation}, i.e., interactions with electrons (red curves), with nucleons via SD interactions (green curves) and with nucleons via SI interactions (blue curves), for a velocity cutoff at $v_c(r) = v_e (r)$ (solid curves) and with a cutoff at $v_c (r) = 0.9 \, v_e (r)$ (dashed curves), and for constant (top panel), $v_{\rm rel}^2$-dependent (bottom-left panel) and $q^2$-dependent (bottom-right panel) cross sections.

For DM masses to the left of the curves, evaporation is very efficient and DM particles evaporate from the Sun. For a DM distribution with a cutoff at the escape velocity, the evaporation mass is slightly larger for interactions with electrons. This is always the case for constant cross sections. However, for $v_{\rm rel}^2$-dependent and $q^2$-dependent cross sections, the transition to the optically thick regime occurs for smaller values of $\sigma_{i, 0}$ for DM-electron interactions than for DM-nucleon interactions, which has to do with the different mass dependence of the mean free path in each case. Whereas in the case of constant cross sections, the evaporation mass for DM-nucleon SD interactions is always slightly smaller than for SI interactions, for $v_{\rm rel}^2$-dependent and $q^2$-dependent cross sections, this only happens in the optically thin regime. Nevertheless, the most interesting and new features appear for DM velocity distributions with a cutoff at $v_c(r) = 0.9 \, v_e(r)$. While this velocity cutoff has a very small impact for DM interactions with nucleons, as could have been anticipated from the previous results, in the case of interactions with electrons, the evaporation mass can be substantially reduced below the GeV range, down to $m_{\rm evap} \sim (0.5-0.6)$~GeV for most values of the cross sections or even to $m_{\rm evap} \sim 0.2$~GeV for some extreme cases, which significantly opens this region of the parameter space, making it potentially testable with neutrino detectors. Therefore, in order to correctly assess the impact of this effect and given the importance of the high-velocity tail in the calculation of the evaporation mass, an accurate evaluation of the equilibrium DM distribution deserves a dedicated analysis. However, this is beyond the scope of this paper.

\section{Neutrino production rates from DM annihilations in the Sun}
\label{sec:rates}

The neutrino production rate from DM annihilations in the Sun is proportional to the annihilation rate of the DM particles accumulated in the Sun and it is given by $\Gamma = A_\odot \, N_\chi^2 / 2$, which after the solution of Eq.~(\ref{eq:evolution}), i.e., Eq.~(\ref{eq:NDM}), results in
\begin{equation}
\label{eq:ratesolution}
\Gamma (\mx, \sigma_{i,0}) = \frac{1}{2} \, C_\odot \, \left(\frac{\tanh(\kappa \, t_\odot/\tau_{\rm eq})}{\kappa + \frac{1}{2} \,E_\odot \, \tau_{\rm eq} \, \tanh(\kappa \, t_\odot/\tau_{\rm eq})}\right)^2 ~.
\end{equation}  
It depends on the DM mass, the scattering cross section and on the annihilation cross section. Therefore, in the limit for which equilibration is attained, $\kappa \, t_\odot \gg \tau_{\rm eq}$, it reads
\begin{equation}
\label{eq:rate}
\Gamma (\mx, \sigma_{i,0}) = \frac{1}{2} \, \frac{C_\odot}{\left(\kappa + \frac{1}{2} \,E_\odot \, \tau_{\rm eq}\right)^2} ~,
\end{equation}  
although in our computations we keep the exact form of Eq.~(\ref{eq:ratesolution}).   

\begin{figure}[t]
	\begin{center}
		\includegraphics[width=0.49\linewidth]{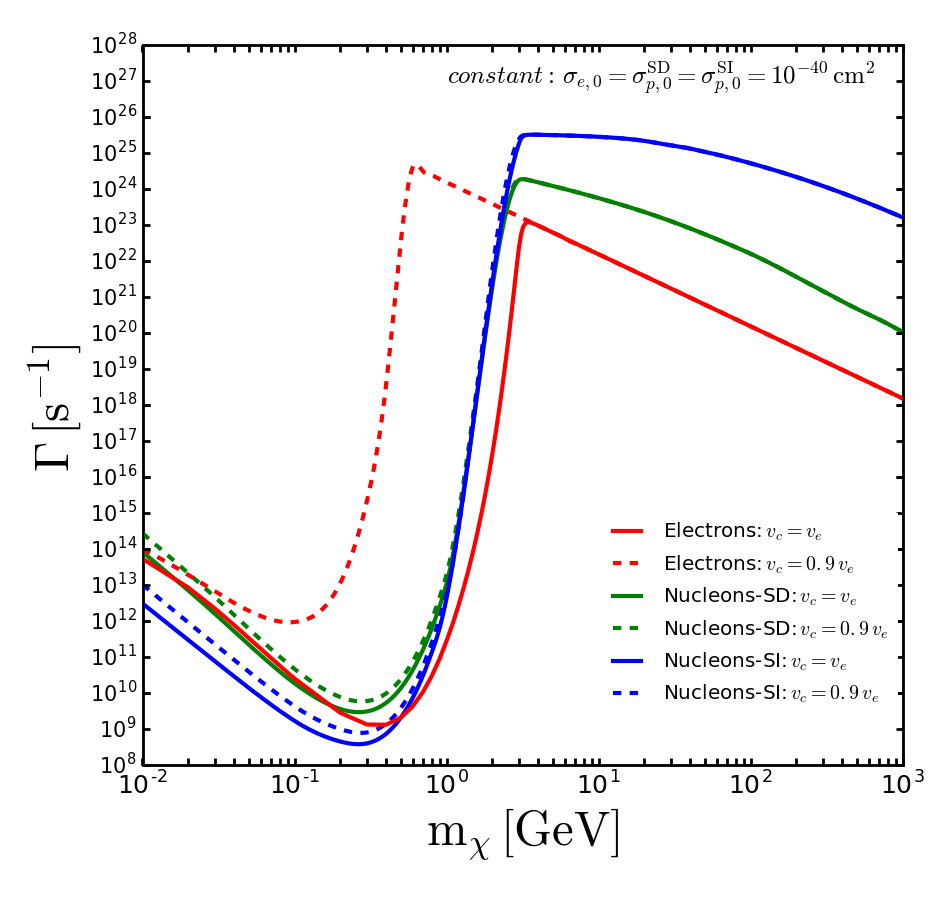} \\		
		\includegraphics[width=0.49\linewidth]{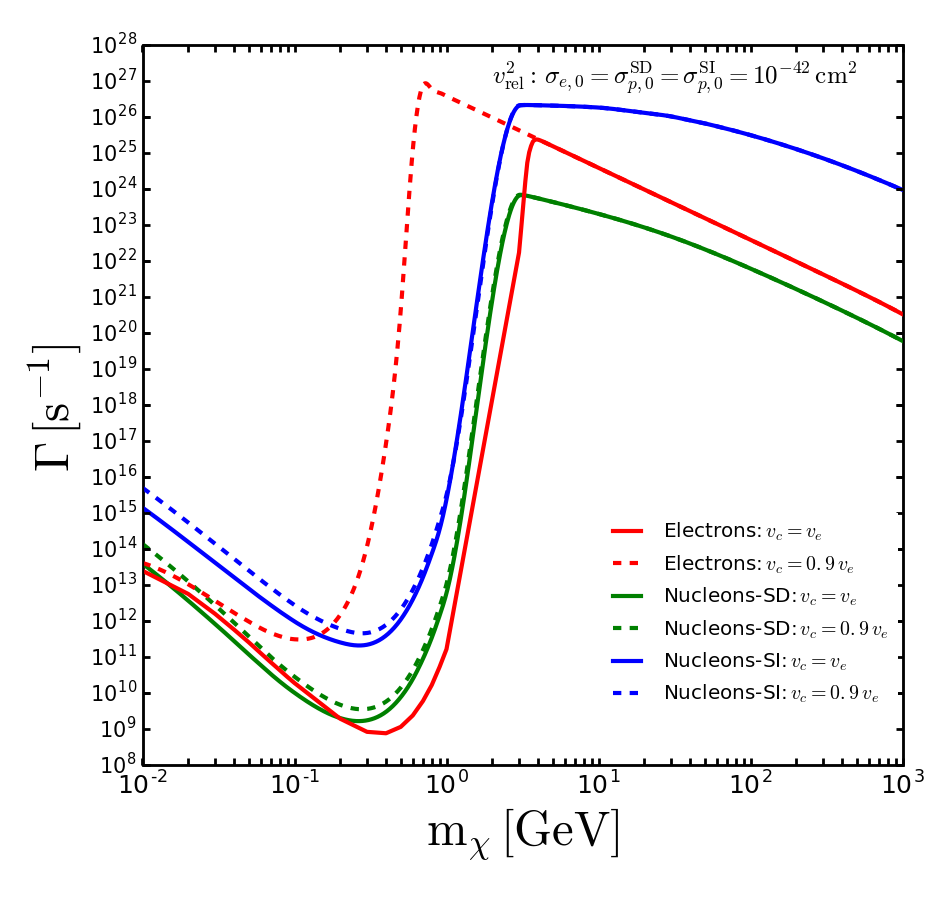}
		\includegraphics[width=0.49\linewidth]{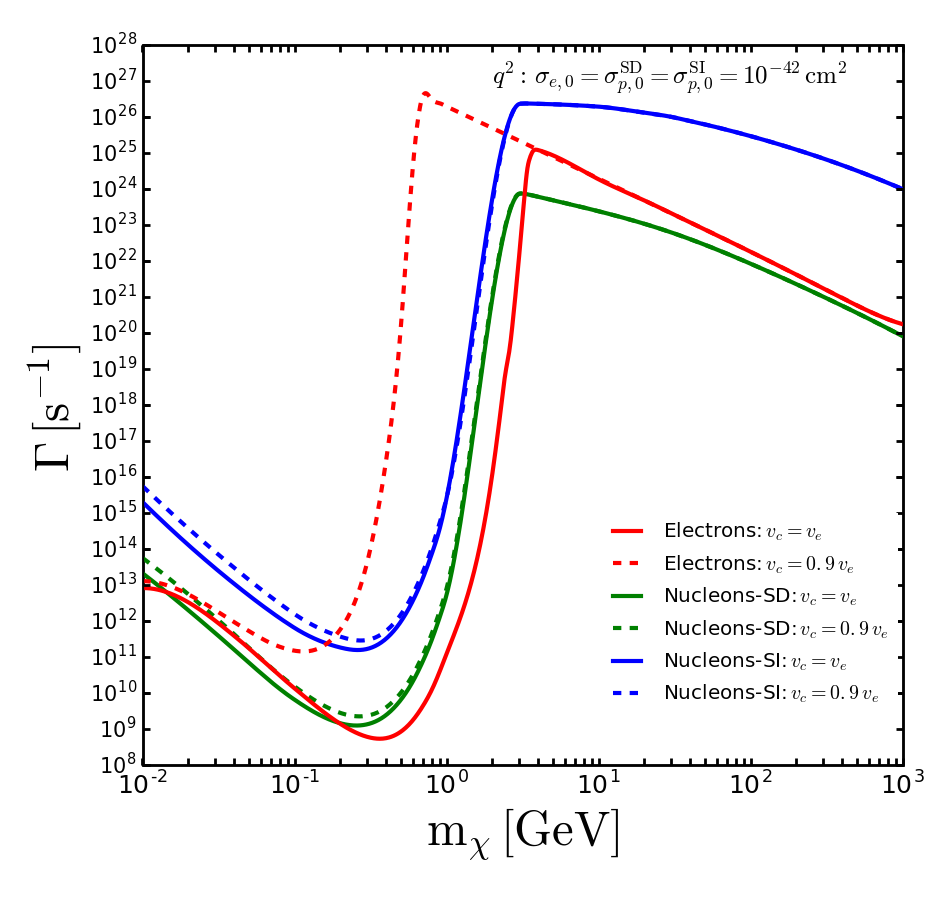}				
	\end{center}
	\caption{\textbf{\textit{Neutrino production rates as a function of the DM mass}}. Same cases as in the left panels of Fig.~\ref{fig:evaporation}. Note the different values of $\sigma_{i,0}$ for the top and bottom panels.}
	\label{fig:rates}
\end{figure}

Once all the ingredients are at hand, we can compare the resulting neutrino rates at production for the case of capture by electrons and by nucleons in the Sun. As already mentioned, even if DM has only tree-level couplings to electrons, loop-induced processes could give rise to interactions with nucleons. Here we compare the relative importance of the different cases considered in this work. In Fig.~\ref{fig:rates} we show the neutrino production rates for the same cases depicted in the left panels of Fig.~\ref{fig:evaporation}, i.e., interactions in the Knudsen limit with $\sigma_{i, 0} = 10^{-40}~{\rm cm}^2$ for constant cross sections (top panel) and $\sigma_{i, 0} = 10^{-42}~{\rm cm}^2$ for $v_{\rm rel}^2$-dependent cross sections (bottom-left panel) and for $q^2$-dependent cross sections (bottom-right panel).

For sufficiently large masses (above the evaporation mass), for which capture and annihilation rates are in equilibrium, the neutrino production rates are proportional to the capture rates, $\Gamma = C_\odot/2$, and hence for all cases, the same behavior described for Fig.~\ref{fig:capture} is obtained. For low masses, equilibrium is attained between capture and evaporation and the neutrino production rate is suppressed by the large evaporation rate, $\Gamma = A_\odot \left(C_\odot/E_\odot\right)^2/2$. Notice that for very low masses, the suppression of the evaporation rate results in the increase of the neutrino production rates, so there is a minimum in the neutrino production rates,\footnote{The rise of the flux at low masses, being much smaller than that at $\mx \gtrsim$~GeV, has no measurable effect.} which lies at $\mx \sim (0.3-0.5)$~GeV for the three types of cross sections we study if the DM velocity distribution extends up to $v_e(r)$. Although the form of the high-velocity tail does not have a strong impact when DM scatters off nucleons, in the case of interactions with electrons, this minimum shifts to $\mx \sim (0.1-0.2)$~GeV for $v_c(r) = 0.9 \, v_e(r)$ and the neutrino production rate is significant for masses $\mx \gtrsim 0.4$~GeV for constant cross sections and $\mx \gtrsim 0.5$~GeV for $v_{\rm rel}^2$-dependent and $q^2$-dependent cross sections. This is to be compared to the minimum DM mass for which there could be a significant neutrino production rate in the case of interactions with nucleons (and with electrons with $v_c(r) = v_e(r)$), the commonly quoted lower limit $\mx \gtrsim 3$~GeV. 

Above the evaporation mass, the neutrino production rates for DM-nucleon SI interactions are the largest for the three types of cross sections (for the normalizations used here). For constant DM-nucleon SD cross sections, these rates are larger than for scatterings off electrons. However, for $v_{\rm rel}^2$-dependent and $q^2$-dependent cross sections, the relative importance gets inverted, being the case of interactions with electrons the most favorable one, because of the larger enhancement due to thermal effects. This illustrates how, for some scenarios, interactions with electrons could give rise to the largest signals in neutrino detectors/telescopes~\cite{Garani:2017}.

\section{Summary}
\label{sec:summary}

The annihilation of DM particles accumulated in the Sun could give rise to a flux of neutrinos with energies of the order of the DM mass from decays of heavy hadrons, gauge bosons and tau leptons~\cite{Silk:1985ax, Krauss:1985aaa, Freese:1985qw, Hagelin:1986gv, Gaisser:1986ha, Srednicki:1986vj, Griest:1986yu, Kamionkowski:1991nj, Bottino:1991dy, Halzen:1991kh, Gandhi:1993ce, Bottino:1994xp, Bergstrom:1996kp, Bergstrom:1998xh, Barger:2001ur, Bertin:2002ky, Hooper:2002gs, Bueno:2004dv,  Cirelli:2005gh, Halzen:2005ar, Mena:2007ty, Lehnert:2007fv, Barger:2007xf, Barger:2007hj, Blennow:2007tw, Liu:2008kz, Hooper:2008cf, Wikstrom:2009kw, Nussinov:2009ft, Menon:2009qj, Buckley:2009kw, Zentner:2009is, Ellis:2009ka, Esmaili:2009ks, Ellis:2011af, Bell:2011sn, Kappl:2011kz,  Agarwalla:2011yy, Chen:2011vda, Kundu:2011ek, Rott:2011fh, Das:2011yr, Kumar:2012uh, Bell:2012dk, Silverwood:2012tp, Blennow:2013pya, Arina:2013jya, Liang:2013dsa, Ibarra:2013eba, Albuquerque:2013xna, Baratella:2013fya, Guo:2013ypa, Ibarra:2014vya, Chen:2014oaa, Blumenthal:2014cwa, Catena:2015iea, Chen:2015uha, Belanger:2015hra, Heisig:2015ira, Danninger:2014xza, Blennow:2015hzp, Murase:2016nwx, Lopes:2016ezf, Baum:2016oow, Allahverdi:2016fvl} and with energies from tens to few hundred MeV from decays of stopped mesons and muons~\cite{Rott:2012qb, Bernal:2012qh, Rott:2015nma, Rott:2016mzs}, which are potentially detectable with neutrino detectors/telescopes~\cite{Desai:2004pq, Desai:2007ra, Abbasi:2009uz, Abbasi:2009vg, Tanaka:2011uf, IceCube:2011aj, Scott:2012mq, Aartsen:2012kia, Adrian-Martinez:2013ayv, Avrorin:2014swy, Choi:2015ara, Aartsen:2016exj, Adrian-Martinez:2016ujo, Aartsen:2016zhm}. The process of capture in astrophysical objects like the Sun is commonly assumed to be due to interactions with nucleons. However, in leptophilic scenarios, in which only couplings to leptons are present at tree level, capture via interactions off electrons could be the only possibility to trap DM particles~\cite{Kopp:2009et}. Moreover, scattering cross sections for DM-electron (and DM-nucleon) interactions, rather than being constant, could depend on the relative velocity ($v_{\rm rel}$) and the scattering angle ($\theta_{\rm cm}$)~\cite{Fan:2010gt, Fitzpatrick:2012ix, Anand:2013yka, Hill:2013hoa, Gresham:2014vja, Panci:2014gga, Catena:2014epa, Gluscevic:2014vga, Catena:2014hla, Gluscevic:2015sqa, Dent:2015zpa, Catena:2015vpa, Kavanagh:2015jma, Catena:2015uha, Gazda:2016mrp}. Indeed, even if loop-induced interactions with nucleons are, in general, possible, scatterings off electrons could be the dominant capture mechanism in certain cases~\cite{Garani:2017}. 

In this work, we have considered DM scatterings off electrons and have studied the different ingredients (capture, annihilation and evaporation) entering the calculation of the neutrino production rates from DM annihilations in the Sun for three type of generic interactions: constant (velocity-independent and isotropic), $v_{\rm rel}^2$-dependent (and isotropic) and $q^2$-dependent cross sections (Section~\ref{sec:cs}). To the best of our knowledge, the possibility of capture via DM-electron interactions had only been considered for the case of constant cross sections and for masses for which evaporation can be neglected ($\mx > 10$~GeV)~\cite{Kopp:2009et}. Here, we have presented detailed analytical and numerical results for the differential scattering rates for three different types of cross sections and generic target particles (Appendix~\ref{app:scatteringrates}), which enter the calculation of the capture (Section~\ref{sec:capture}) and evaporation (Section~\ref{sec:evaporation}) rates; we have presented refinements in the calculation of the temperature in the isothermal approximation (Appendix~\ref{app:temperature}), which slightly modify the DM distribution in the Sun with respect to the standard case (Section~\ref{sec:distribution}) and thus, could affect the annihilation and evaporation rates; and we have also computed the mean free path of DM particles for each type of cross section in generic terms (Appendix~\ref{app:mfp}), which is relevant for the computation of the DM distribution in the Sun and for the suppression factor appearing in the evaporation rate (Appendix~\ref{app:supp}). Moreover, we have also investigated the effects of the truncation of the DM distribution at a velocity smaller than the escape velocity.

Given that for leptophilic DM models loop-induced interactions with nucleons would, in general, be present, we compare all our results on DM capture by electrons to those obtained for DM capture by nuclei. All our computations of capture rates take into account thermal effects (left panels of Fig.~\ref{fig:capture}), which are very important for interactions off electrons~\cite{Kopp:2009et}, but are usually neglected for scatterings off nucleons (here presented for the first time for $v_{\rm rel}^2$-dependent and $q^2$-dependent cross sections). Indeed, for DM-electron scatterings the enhancement of the capture rate with respect to the zero-temperature limit could be up to three orders of magnitude for $v_{\rm rel}^2$-dependent and $q^2$-dependent cross sections. On the other hand, we confirm the well-known fact that, in the case of constant cross sections, thermal effects are negligible for capture by nucleons, but we note that for $v_{\rm rel}^2$-dependent and $q^2$-dependent cross sections, these effects could be significant (see the right panels of Fig.~\ref{fig:capture}). 

For low DM masses (typically of a few GeV), evaporation is a very effective process to reduce the number of DM particles accumulated in the Sun~\cite{Steigman:1997vs,Spergel:1984re, Faulkner:1985rm, Krauss:1985aaa, Gilliland:1986, Nauenberg:1986em, Griest:1986yu, Gould:1987ju, Gould:1990}, so we have also evaluated the evaporation mass as a function of the scattering cross section, i.e., the minimum mass  for which DM could remained trapped in the Sun (Section~\ref{sec:evapmass}). Whereas for the case of nucleons the evaporation mass does not depend much on the cutoff (comparing $v_c(r) = v_e(r)$ and $v_c(r) = 0.9 \, v_e(r)$) of the DM velocity distribution (Fig.~\ref{fig:evapmass}), for interactions with electrons, the presence of a cutoff $v_c(r) < v_e(r)$ could have important implications, shifting the evaporation mass below the GeV range, down to few-hundred MeV, and opening up a new region in the parameter space suitable to be tested in the future. A definite answer regarding this possibility would require the use of the correct DM distribution and thus, solving the collisional Boltzmann equation, which is beyond the scope of this paper.

Finally, we have compared the neutrino rates at production resulting from ($s$-wave) annihilations of DM particles after being captured either by solar electrons or nuclei for constant, $v_{\rm rel}^2$-dependent and $q^2$-dependent scattering cross sections (Fig.~\ref{fig:rates}). We have found that, for the normalizations of the cross sections considered in this work, capture by electrons would result in neutrino rates about two orders of magnitude smaller than those obtained in the case of DM-nucleon SD interactions for constant cross sections, whereas in the case of $v_{\rm rel}^2$-dependent and $q^2$-dependent cross sections, interactions off electrons result in a larger neutrino production rate. For the three type of cross sections, the most efficient process  is via DM-nucleon SI interactions, although stronger limits exist in this case~\cite{Felizardo:2011uw, Abe:2015eos, Agnese:2015nto, Amole:2016pye, Aprile:2016wwo, Tan:2016zwf, Akerib:2016vxi, Aprile:2016swn, Aprile:2017yea}.

So far, there is no conclusive evidence of the existence of DM, other than from its gravitational interactions. Therefore, investigating different and complementary techniques to search for DM is of crucial importance. Here, we have studied  one of the existing strategies to indirectly detect DM, which is in turn complementary to DM direct searches. Indeed, the phenomenological approach discussed in much detail in this work represents the first step in evaluating the relevance of DM capture in leptophilic scenarios and of their potential signals at neutrino detectors/telescopes~\cite{Garani:2017}.

\section*{Acknowledgments}
SPR dedicates this work to the memory of his friend and collaborator Haim Goldberg. \linebreak RG is supported by the German Research Foundation through TRR33 ``The Dark Universe'', the Helmholtz Alliance for Astroparticle Physics and the Bonn-Cologne graduate school. SPR is supported by a Ram\'on y Cajal contract, by the Spanish MINECO under grants FPA2014-54459-P and SEV-2014-0398, by the Generalitat Valenciana under grant PROMETEOII/2014/049 and by the European Union's Horizon 2020 research and innovation program under the Marie Sk\l odowska-Curie grant agreements No. 690575 and 674896. SPR is also partially supported by the Portuguese FCT through the CFTP-FCT Unit 777 (PEst-OE/FIS/UI0777/2013).

\appendix

\section{Differential scattering rates}
\label{app:scatteringrates}

We consider DM particles with mass $\mx$ interacting with a thermal distribution of targets with mass $m_i$, such that we can work on the non-relativistic limit. Assuming the target particles follow a Maxwell-Boltzmann velocity distribution, $f_i(\boldsymbol{u},r)$, with density $n_i(r)$ and temperature $T_\odot(r)$, the differential rate at which a DM particle with velocity $w$ scatters off a target $i$ with velocity $u$ and relative angle $\theta$, in the laboratory frame, to a final velocity $v$ is given by
\begin{eqnarray}
\label{eq:scatteringrate}
R_i (w \to v) & = & \int n_i(r) \, \frac{\dd \sigma_i}{\dd v} \, |\boldsymbol{w} - \boldsymbol{u}| \, f_i(\boldsymbol{u},r) \, \dd^3 \boldsymbol{u} \nonumber \\
& = & \frac{2}{\sqrt{\pi}} \, \frac{n_i(r)}{u_i^3(r)} \, \int_0^\infty \dd u \, u^2 \, \int_{-1}^1 \dd \cos\theta \, \frac{\dd \sigma_i}{\dd v} \, |\boldsymbol{w} - \boldsymbol{u}| \, e^{-u^2/u_i^2(r)}  ~, 
\end{eqnarray}
where $\dd \sigma_i/\dd v$ is the differential scattering cross section, 
\begin{equation}
\label{eq:ui}
u_i(r) \equiv \sqrt{\frac{2 \, T_\odot(r)}{m_i}} ~, 
\end{equation}%
is the most probable speed of the target particles at position $r$, and the relative velocity between the DM and target particles is given by
\begin{equation}
\label{eq:vrel}
|\boldsymbol{w} - \boldsymbol{u}| = \sqrt{w^2 + u^2 - 2 \, w \, u \, \cos{\theta}} ~.
\end{equation} 

Now, using the notation of Refs.~\cite{Gould:1987ju, Gould:1989hm}, we can write the above expression in terms of the velocity of the center of mass, $\boldsymbol{s}$, and the velocity of the incoming DM particle in the center-of-mass frame, $\boldsymbol{t}$,
\begin{eqnarray}
\label{eq:st}
(1 + \mu_i) \, \boldsymbol{s} & = & \boldsymbol{u} + \mu_i \, \boldsymbol{w} ~, \\
(1 + \mu_i) \, \boldsymbol{t} & = & \boldsymbol{w} - \, \boldsymbol{u} ~,
\end{eqnarray}
where we have defined 
\begin{equation}
\label{eq:mu}
\mu_i \equiv \frac{\mx}{m_i} ~.
\end{equation}
If we substitute $u$ and $\cos \theta$ by $s$ and $t$, Eq.~(\ref{eq:scatteringrate}) can be expressed as
\begin{equation}
\label{eq:rate-st}
R_i (w \to v) = \frac{32}{\sqrt{\pi}} \, \frac{\mu_{i,+}^4}{w \, u_i^3(r)} \, n_i(r) \, \int_0^\infty \dd s  \int_0^\infty \dd t  \, \frac{\dd \sigma_i}{\dd v} \, s \, t^2 \, e^{-u^2/u_i^2(r)} \, \Theta(w - |s - t|) \, \Theta(s + t - w)
\end{equation}
where 
\begin{equation}
\label{eq:mudef}
\mu_{i,\pm} \equiv \frac{\mu_i \pm 1}{2} ~, \hspace{1cm} u^2 = 2 \, \mu_i \, \mu_{i,+} \, t^2 + 2 \, \mu_{i,+} \, s^2 - \mu_i \, w^2 ~.  
\end{equation}
On the other hand, the differential cross section $\dd \sigma_i/\dd v$ can be written as
\begin{equation}
\label{eq:dsdv}
\frac{\dd \sigma_i}{\dd v} = \int_0^{2 \pi} \frac{1}{2 \pi} \, \frac{\dd \sigma_i}{\dd \cos \theta_{\rm cm}} \frac{\dd \cos \theta_{st'}}{\dd v} \, \Theta(1-|\cos \theta_{st'}|) \, \dd \phi_{st'} ~, 
\end{equation}
where $\theta_{\rm cm}$ is the center-of-mass angle between the velocity of the outgoing DM particle, $\boldsymbol{t'}$, and the velocity of the incoming DM particle, $\boldsymbol{t}$,
\begin{equation}
\label{eq:thetacm}
\cos \theta_{\rm cm} = \cos \theta_{st} \, \cos \theta_{st'} + \sin \theta_{st} \, \sin \theta_{st'} \, \cos \phi_{st'} ~,
\end{equation}
and $\theta_{st'}$ and $\phi_{st'}$ are the center-of-mass angles of the outgoing DM particle with respect to $\boldsymbol{s}$ and $\theta_{st}$ is the center-of-mass zenith angle of the incoming DM particle with respect to $\boldsymbol{s}$ (see, e.g., Fig.~1 in Ref.~\cite{Vincent:2013lua}), 
\begin{eqnarray}
\label{eq:cosst}
\cos \theta_{st} & = &  \frac{w^2 - s^2 - t^2}{2 \, s \, t} ~, \\[2ex]
\cos \theta_{st'} & = & \frac{v^2 - s^2 - t^2}{2 \, s \, t} ~.
\end{eqnarray}
Therefore, we can express the differential DM scattering rate off target $i$, Eq.~(\ref{eq:rate-st}), as
\begin{equation}
\label{eq:rate-stcm}
R_i (w \to v) = \frac{16}{\sqrt{\pi^3}} \, \frac{\mu_{i,+}^4}{u_i^3(r)} \, \frac{v}{w} \, n_i(r) \, \int_0^\infty \dd t \int_0^\infty \dd s  \, t \, e^{-u^2/u_i^2(r)} \, H(s,t,w,v) \, \int_0^{2 \pi} \dd \phi_{st'} \, \frac{\dd \sigma_i}{\dd \cos \theta_{\rm cm}}  ~,
\end{equation}
with
\begin{equation}
\label{eq:H}
H(s,t,w,v) \equiv \Theta(w - |s - t|) \, \Theta(s + t - w) \, \Theta(v - |s - t|) \, \Theta(s + t - v) ~.
\end{equation}

If one performs the $s$ integral first, this product of Heaviside functions translates into the following integration limits for $s$ and $t$:
\begin{eqnarray}
\label{eq:intlim}
(\textrm{for} \, \, v < w) \, H^-(s,t,w,v):  & &  \\
& & \frac{w-v}{2} \le t \le \frac{v + w}{2}  ~, \hspace{1cm} w - t \le s \le v + t ~, \nonumber \\[1ex]
& & \frac{v + w}{2} \le t \le \infty  ~, \hspace{1.7cm} t - v \le s \le v + t ~; \nonumber \\
(\textrm{for} \, \, v > w) \, H^+(s,t,w,v):  & & \\
& & \frac{v - w}{2} \le t \le \frac{v + w}{2}  ~, \hspace{1cm} v - t \le s \le w + t ~, \nonumber\\[1ex]
& & \frac{v + w}{2} \le t \le \infty  ~, \hspace{1.7cm} t - w \le s \le w + t ~. \nonumber 
\end{eqnarray}

Finally, the full differential scattering rate is obtained by summing over all targets $i$, i.e, $R(w \to v) = \sum_i R_i(w \to v)$. In what follows we also use these definitions~\cite{Gould:1987ju}:
\begin{eqnarray}
\label{eq:ab}
\chi(a,b) & \equiv & \int_a^b e^{-y^2} \, \dd y ~, \\[1ex]
\alpha_{\pm} & \equiv & \frac{\mu_{i,+} \, v \pm \mu_{i,-} \, w}{u_i(r)} ~, \\[1ex]
\beta_{\pm} & \equiv & \frac{\mu_{i,-} \, v \pm \mu_{i,+} \, w}{u_i(r)} ~.
\end{eqnarray}

\subsection{Constant cross section: velocity-independent and isotropic}

We first consider velocity-independent and isotropic cross sections, which is the case usually studied in the literature,
\begin{equation}
\label{eq:dss}
\frac{\dd \sigma_{i,\rm const} (v_{\rm rel}, \cos \theta_{\rm cm})}{\dd \cos{\theta_{\rm cm}}} = \frac{\sigma_{i,0}}{2} ~,
\end{equation}
so that $\sigma_{i,\rm const} = \sigma_{i,0}$. Therefore, Eq.~(\ref{eq:rate-stcm}) reads
\begin{equation}
\label{eq:rate-s}
R_{i, \rm const}^{\pm} (w \to v) = \frac{16}{\sqrt{\pi}} \, \frac{\mu_{i,+}^4}{u_i^3(r)} \, \frac{v}{w} \, n_i(r) \, \sigma_{i,0} \, \int_0^\infty \dd t \int_0^\infty \dd s  \, t \, e^{-u^2/u_i^2(r)} \, H^{\pm}(s,t,w,v)  ~.
\end{equation}
Using the conditions $v>w$ for $R_{i, \rm const}^+ (w \to v)$, and $v<w$ for $R_{i, \rm const}^- (w \to v)$, the integrals in Eq.~(\ref{eq:rate-s}) can be performed analytically~\cite{Gould:1987ju},
\begin{equation}
\label{eq:ints}
\int_0^\infty \dd t \int_0^\infty \dd s  \, t \, e^{-u^2/u_i^2(r)} \, H^{\pm}(s,t,w,v) = \frac{u_i^3(r)}{8 \, \mu_i \, \mu_{i,+}^2} \, \left[ \chi(\pm \alpha_-,\alpha_+) + \chi(\pm \beta_-,\beta_+) \, e^{\mu_i  \left(w^2-v^2\right)/u_i^2(r)} \right] .
\end{equation}
Finally, the rates $R_{\rm const}^{\pm} (w \to v)$ are given by~\cite{Gould:1987ju}
\begin{equation}
\label{eq:rates}
R_{\rm const}^{\pm} (w \to v) = \sum_i \frac{2}{\sqrt{\pi}} \, \frac{\mu_{i,+}^2}{\mu_i} \, \frac{v}{w} \, n_i(r) \, \sigma_{i,0} \, \left[ \chi(\pm \alpha_-,\alpha_+) + \chi(\pm \beta_-,\beta_+) \, e^{\mu_i  \left(w^2-v^2\right)/u_i^2(r)} \right] ~. 
\end{equation}

\subsection{Velocity-dependent cross section}

We consider now the case of velocity-dependent and isotropic cross sections, i.e., 
\begin{equation}
\label{eq:dsvn}
\frac{\dd \sigma_{i,v_{\rm rel}^n} (v_{\rm rel}, \cos \theta_{\rm cm})}{\dd \cos{\theta_{\rm cm}}} = \frac{\sigma_{i,0}}{2} \, \left(\frac{v_{\rm rel}}{v_0}\right)^n ~,
\end{equation}
so that $\sigma_{i,v_{\rm rel}^n} = \sigma_{i,0} \, (v_{\rm rel}/v_0)^n$, and where $v_{\rm rel}$ is the relative velocity between the DM particle and target $i$,
\begin{equation}
\label{eq:vreldef}
v_{\rm rel} \equiv |\boldsymbol{w} - \boldsymbol{u}| = (1 + \mu_i) \, t = 2 \, \mu_{i,+} \, t~, 
\end{equation} 
and $v_0$ is an arbitrary reference velocity, which we set to $v_0 = 220$~km/s. Thus, for this particular case, the differential scattering rates read
\begin{equation}
\label{eq:ratevn}
R_{i, v_{\rm rel}^n}^{\pm} (w \to v) = \frac{2^{4+n}}{\sqrt{\pi}} \, \frac{\mu_{i,+}^{4+n}}{u_i^3(r)} \, \frac{v}{w} \, n_i(r) \, \frac{\sigma_{i,0}}{v_0^n} \, \int_0^\infty \dd t  \int_0^\infty \dd s \, t^{1+n} \, e^{-u^2/u_i^2(r)} \, H^{\pm}(s,t,w,v) .
\end{equation}
For the case of $n=2$, using the conditions $v>w$ for $R_{i,v_{\rm rel}^2}^+ (w \to v)$, and $v<w$ for $R_{i,v_{\rm rel}^2}^- (w \to v)$, the integrals in Eq.~(\ref{eq:ratevn}) can be performed analytically,
\begin{eqnarray}
\label{eq:intv2}
&& \int_0^\infty \dd t \int_0^\infty \dd s \, t^3 \, 
e^{-u^2/u_i^2(r)} \, H^{\pm}(s,t,w,v)  =   \nonumber \\
&& \hspace{5mm} \frac{u_i^5(r)}{32 \, \mu_i \, \mu_{i,+}^4}  \, \left[\left(\mu_{i,+} + \frac{1}{2}\right) \, \left(\pm \frac{v-w}{u_i(r)} \, e^{-\alpha _-^2} - \frac{v+w}{u_i(r)} \, e^{-\alpha _+^2}\right)  \right. \\ 
&&  \hspace{5mm} \left. + \, \left(\frac{w^2}{u_i^2(r)} + \frac{3}{2} + \frac{1}{\mu_i} \right) \, \chi(\pm \alpha_-, \alpha_+)   + \,  \left(\frac{v^2}{u_i^2(r)} + \frac{3}{2} + \frac{1}{\mu_i} \right) \, \chi(\pm \beta_-,\beta_+) \, e^{\mu_i  \left(w^2-v^2\right)/u_i^2(r)} \right] ~, \nonumber
\end{eqnarray}
Finally, the rates $R_{v_{\rm rel}^2}^{\pm} (w \to v)$ are given by
\begin{eqnarray}
\label{eq:ratev2}
R_{v_{\rm rel}^2}^{\pm} (w \to v) & = & \sum_i \frac{2}{\sqrt{\pi}} \, \frac{\mu_{i,+}^2}{\mu_i} \, \frac{v}{w} \, n_i(r) \, \sigma_{i,0}  \, \left(\frac{u_i(r)}{v_0}\right)^2 \left[\left(\mu_{i,+} + \frac{1}{2}\right) \, \left(\pm \frac{v-w}{u_i(r)} \, e^{-\alpha _-^2} - \frac{v+w}{u_i(r)} \, e^{-\alpha _+^2} \right) \right. \nonumber \\
& & \left. +\, \left(\frac{w^2}{u_i^2(r)} + \frac{3}{2} + \frac{1}{\mu_i} \right) \, \chi(\pm \alpha_-, \alpha_+) \right. \nonumber \\
& & \left. + \left(\frac{v^2}{u_i^2(r)} + \frac{3}{2} + \frac{1}{\mu_i} \right) \, \chi(\pm \beta_-,\beta_+) \,  e^{\mu_i  \left(w ^2-v^2\right)/u_i^2(r)} \right] ~. \nonumber \\
\end{eqnarray}

\subsection{Momentum-dependent cross section}

Next, we consider the case of momentum-dependent cross sections,
\begin{equation}
\label{eq:dsqn}
\frac{\dd \sigma_{i,q^n} (v_{\rm rel}, \cos \theta_{\rm cm})}{\dd \cos{\theta_{\rm cm}}} = \frac{\sigma_{i,0}}{2} \, H_{i,n} \, \left(\frac{q}{q_0}\right)^n ~,
\end{equation}
where
\begin{equation}
\label{eq:Hn}
H_{i,n} = \frac{2^{n/2+1} \, \mu_{i,+}^n}{\int_{-1}^1 \, \left(1 - \cos\theta_{\rm cm}\right)^{n/2} \, \dd \cos\theta_{\rm cm}} = \left( 1 + \frac{n}{2}\right) \, \mu_{i,+}^n \, 
\end{equation}
is a normalization chosen so that the total momentum-dependent and velocity-dependent cross sections are equal, i.e., $\sigma_{i,q^n} = \sigma_{i,v_{\rm rel}^n}$, if the arbitrary momentum is defined as $q_0 = \mx \, v_0$ and $\sigma_{i,0}$ is the same in both cases. In the non-relativistic limit, the 3-momentum transfer is given by
\begin{equation}
q^2 = m_{\chi}^2 \, |\boldsymbol{w} - \boldsymbol{v}|^2 = 2 \, \mx^2 \, t^2 \, (1 - \cos \theta_{\rm cm}) = \frac{\mx^2}{2 \, \mu_{i,+}^2} \, v_{\rm rel}^2 \, (1 - \cos \theta_{\rm cm}) ~.
\end{equation}

Then, the differential scattering rates read
\begin{eqnarray}
\label{eq:rateqn}
R_{i,q^n}^{\pm} (w \to v) & = & \frac{2^{4 + n/2}}{\pi^{3/2}} \, \frac{\mu_{i,+}^{4+n}}{u_i^3(r)} \, \frac{v}{w} \, n_i(r) \, \frac{\sigma_{i,0}}{v_0^n} \nonumber \\
& & \times \int_0^\infty \dd t \int_0^\infty \dd s  \,  t^{1+n} \, e^{-u^2/u_i^2(r)} \, H^{\pm}(s,t,w,v) \int_0^{2 \pi} \dd \phi_{st'} \, \left(1- \cos \theta_{\rm cm}\right)^{n/2}  ~. \nonumber \\
\end{eqnarray}
For the case $n=2$, using the conditions $v>w$ for $R_{i,q^2}^+ (w \to v)$, and $w>v$ for $R_{i,q^2}^- (w \to v)$, the integrals in Eq.~(\ref{eq:rateqn}) can be performed analytically. After computing the $\phi_{st'}$ integral, one obtains

\begin{eqnarray}
\label{eq:intq2}
& & 2 \pi \int_0^\infty \dd t \int_0^\infty \dd s \, \left(1 - \frac{(w^2 -s^2 -t^2)(v^2 -s^2 -t^2)}{4 \, s^2 \, t^2}\right) \, t^3 \, e^{-u^2/u_i^2(r)} \, H^{\pm}(s,t,w,v) =   \nonumber \\ 
& & \frac{\pi}{4} \, \frac{u_i^5(r)}{\mu_i^2 \, \mu_{i,+}^2} \, \left[\pm \frac{v-w}{u_i(r)} \, e^{-\alpha_-^2} -  \frac{w + v}{u_i(r)} \, e^{-\alpha_+^2} \right. \\ 
& & \left . +  
\left(\frac{1}{2}  \, \frac{w^2 -v^2}{u_i^2(r)} + \frac{1}{\mu_i} \right)  \, \chi(\pm \alpha_-,\alpha_+) + \left(\frac{1}{2} \, \frac{v^2 - w^2}{u_i^2(r)} + \frac{1}{\mu_i} \right) \, \chi(\pm \beta_-,\beta_+) \, e^{\mu_i \, \left(w^2-v^2\right)/u_i^2(r)} \right] ~.\nonumber
\end{eqnarray}

Finally, the rates $R_{q^2}^{\pm} (w \to v)$ are given by
\begin{eqnarray}
\label{eq:rateq2}
R_{q^2}^{\pm}(w \to v) & = & \sum_i \frac{8}{\sqrt{\pi}} \, \frac{\mu_{i,+}^4}{\mu_i^2} \,\frac{v}{w} \, n_i(r) \,  \sigma_{i,0} \, \left(\frac{u_i(r)}{v_0}\right)^2 \, \left[\pm \frac{v-w}{u_i(r)} \, e^{-\alpha_-^2} -  \frac{w + v}{u_i(r)} \, e^{-\alpha_+^2} \right.  \nonumber \\ 
& & \left . +  
\left(\frac{1}{2}  \, \frac{w^2 -v^2}{u_i^2(r)} + \frac{1}{\mu_i} \right)  \, \chi(\pm \alpha_-,\alpha_+) \right. \nonumber \\
& & \left. + \left(\frac{1}{2} \, \frac{v^2 - w^2}{u_i^2(r)} + \frac{1}{\mu_i} \right) \, \chi(\pm \beta_-,\beta_+) \, e^{\mu_i \, \left(w^2-v^2\right)/u_i^2(r)} \right] ~. 
\end{eqnarray}

\section{Dark matter temperature: the Knudsen limit}
\label{app:temperature}

Although a Maxwell-Boltzmann distribution, Eq.~(\ref{eq:DMdistiso}), is not an exact solution of the collisional Boltzmann equation in the optically thin regime (the Knudsen limit)~\cite{Gould:1987ju, Gould:1989ez, Liang:2016yjf}, one can obtain a solution for the isothermal approximation by requiring the distribution to satisfy its first energy moment and solving for $\tx$. In the case of the collisionless Boltzmann equation, for a steady state equilibrium distribution without evaporation, the energy moment equation implies no net flow of energy~\cite{Spergel:1984re}, i.e.,\footnote{Note that the condition only on the first moment starts to fail for velocities $w^2 > 2 \, \tx/\mx$, i.e., for $\mx \gtrsim {\cal O}(0.1)$~GeV, although we will also use it for slightly lower masses.}
\begin{equation}
\label{eq:nonetflow0}
\sum_i \int_0^{R_\odot} \epsilon_i(r, \tx, v_c) \, 4 \pi r^2 \, \dd r = 0 ~,
\end{equation}
where
\begin{eqnarray}
\label{eq:eps}
\epsilon_i(r, \tx, v_c) \equiv  \int \dd^3 \boldsymbol{w} \, n_{\chi,{\rm iso}}(r) \, f_{\chi, {\rm iso}}(\boldsymbol{w},\boldsymbol{r}) \int \dd^3 \boldsymbol{u} \, n_i(r) \, f_i(\boldsymbol{u},r)  \, \sigma_{i,0} \, |\boldsymbol{w} - \boldsymbol{u}| \, \langle \Delta E_i \rangle
\end{eqnarray}
is the energy transfer per unit volume and time, and where the velocity distribution functions are given in Eqs.~(\ref{eq:fv}) and~(\ref{eq:DMdistiso}), which we reproduce here (in the DM case, for the isothermal approximation),
\begin{eqnarray}
\label{eq:fvapp}
f_i(\boldsymbol{u},r) & = & \frac{1}{\sqrt{\pi^3} \, u_i^3(r)} \, e^{- u^2/u_i^2(r)} ~, \\
\label{eq:DMdistisoapp}
f_{\chi, {\rm iso}}(\boldsymbol{w},r) & = & \frac{e^{- w^2/\vx^2} \, \, \Theta(v_c(r) - w)}{\sqrt{\pi^3} \, \vx^3 \, \left(\text{Erf}\left(\frac{v_c(r)}{\vx}\right) - \frac{2}{\sqrt{\pi}} \, \frac{v_c(r)}{\vx} \, e^{- v_c^2(r)/\vx^2}\right)} ~,
\end{eqnarray}
where $u_i(r)$ is defined in Eq.~(\ref{eq:ui}), 
\begin{equation}
\label{eq:vx}
\vx \equiv \sqrt{\frac{2 \, \tx}{\mx}} ~,
\end{equation}
$v_c(r)$ is the (position-dependent) cutoff velocity of the DM distribution and
\begin{equation}
\label{eq:intDEdef}
\langle \Delta E_i \rangle = \int_{-1}^1 \frac{\dd \cos\theta_{\rm cm}}{\sigma_{i,0}} \, \frac{\dd \sigma_i}{\dd \cos\theta_{\rm cm}} \, \Delta E_i(\boldsymbol{w},\boldsymbol{u},\cos\theta_{\rm cm})
\end{equation}
with $\Delta E_i$ being the energy transferred to the DM particle after one collision,
\begin{equation}
\label{eq:DE}
\Delta E_i \equiv \frac{\mx}{2} \, \left(v^2 - w^2\right) ~.
\end{equation}
In terms of the incoming velocities and the scattering angle in the laboratory frame, $\theta$, and in terms of the scattering angle in the center-of-mass frame, $\theta_{\rm cm}$, Eq.~(\ref{eq:DE}) reads~\cite{Spergel:1984re}
\begin{equation}
\label{eq:DEcm}
\Delta E_i(\boldsymbol{w},\boldsymbol{u},\cos\theta_{\rm cm}) = \frac{\mx}{4 \, \mu_{i,+}^2}  \, (1 - \cos\theta_{\rm cm}) \, \left(u^2 - \mu_i \, w^2 + 2 \, \mu_{i,-} \, w \, u \, \cos\theta \right) ~.
\end{equation}

More explicitly, in terms of the gravitational potential $\phi(r)$ (see Eq.~(\ref{eq:nDMr})), Eq.~(\ref{eq:nonetflow0}) can be written as
\begin{eqnarray}
\label{eq:nonetflow2}
\sum_i \int_0^{R_\odot} \dd r \, r^2 \, n_i(r) \, e^{\frac{- \mx \, \phi(r)}{ \tx}} \int_0^{v_c(r)} \dd w \, w^2 \, f_{\chi, {\rm iso}}(\boldsymbol{w},r) & & \nonumber \\
\times \int_0^\infty \dd u \, u^2 \, f_i(\boldsymbol{u},r) \int_{-1}^1 \dd \cos\theta \, \sigma_{i,0} \, |\boldsymbol{w} - \boldsymbol{u}| \, \langle \Delta E_i \rangle & = & 0 ~.
\end{eqnarray}

\subsection{Constant cross section: velocity-independent and isotropic}

For the usually considered case, constant (velocity-independent and isotropic) scattering cross section, the expression for the $\cos\theta$ integral reads
\begin{eqnarray}
\label{eq:DEs}
\int_{-1}^1 \dd \cos\theta \,  |\boldsymbol{w} - \boldsymbol{u}|  \, \langle \Delta E_i \rangle & = & \frac{\mx}{60 \, \mu_{i,+}^2} \, \frac{1}{w \, u} \, \bigg[ \bigg(\mu_i \, (4 \, w + u) + (w + 4 \, u)\bigg) \, (w-u)^3 \, |w-u|  \nonumber \\
& & \hspace{2cm} - \bigg((4 \, \mu_i + 1) \, w - (\mu_i + 4)\bigg) \, (w + u)^4)\bigg] ~.
\end{eqnarray} 

One can also perform analytically the velocity integrals. If there is no cutoff in the DM velocity distribution, i.e., $v_c(r) = \infty$, the equation to be solved to obtain the solution for $\tx$ can be written as~\cite{Spergel:1984re}
\begin{equation}
\label{eq:tx-s}
\sum_i \int_0^{R_\odot} \sigma_
{i,0} \,  \, n_i(r)\,  \frac{\mx m_i}{(m_i + \mx)^2} \, \left(\frac{m_i \tx + \mx T_\odot(r)}{\mx m_i}\right)^{1/2} \, \left( T_\odot(r) - \tx  \right) \,  e^{\frac{- \mx \, \phi(r)}{ \tx}} \, r^2 \, \dd r = 0 ~.
\end{equation}
On the other hand, if there is a (position-dependent) cutoff on the DM velocity distribution at $v_c(r)$ and we define $T_c(r) \equiv \mx v_c^2(r)/2$ , then
\begin{eqnarray}
\label{eq:epsT-s}
\epsilon_{i, \rm const}(r, \tx, T_c) & = & \frac{N_\chi \, e^{- \mx \,  \phi(r)/\tx}}{\left( \text{Erf}\left(\sqrt{\frac{T_c(r)}{\tx}}\right) - \sqrt{\frac{4 \, T_c(r)}{\pi \, \tx}} \, e^{-T_c(r)/\tx} \right) \, 4 \pi \int_0^{R_\odot} \dd r \, r^2  \, e^{-\mx \phi(r)/\tx}} \nonumber \\ 
& & \times  \frac{2}{\sqrt{\pi}} \, \frac{\mu_i}{\mu_{i,+}^2} \, \sqrt{\frac{2}{\mx}} \, n_i(r) \, \sigma_{i,0} \, \bigg\{  \nonumber \\
& & \text{Erf}\left(\sqrt{\frac{T_c(r)}{\mu_i \, T_\odot(r)}}\right) \, \frac{e^{-T_c(r)/\tx}}{\sqrt{\tx}} \left[
\frac{1}{2} \, \left(\frac{\mu_i}{2} \, T_\odot(r) + T_c(r)\right)^2  \right. \nonumber \\
& & \left. \left. - \, \left(\frac{\mu_i}{2} \, T_\odot(r)\right)^2 - \left(\frac{\mu_i}{2} \, T_\odot(r) + \tx + T_c(r)\right) \, \left(T_\odot(r) - \tx\right) \right. \bigg] \right. \nonumber \\
& & + \, \sqrt{\frac{\mu_i \, T_c(r) \, T_\odot(r)}{\pi \, \tx}} \, e^{-\left(\frac{T_c(r)}{\tx} + \frac{T_c(r)}{\mu_i T_\odot(r)}\right)}  \, \left[ \frac{1}{2} \, \left(\frac{\mu_i}{2} \, T_\odot(r) + T_c(r)\right)  -  \left(T_\odot(r) - \tx\right) \right] \nonumber \\
& & + \, \left(\mu_i \, T_\odot(r) + \tx\right)^{1/2} \, \left(T_\odot(r) - \tx\right) \, \text{Erf}\left(\sqrt{\frac{T_c(r)}{\tx} + \frac{T_c(r)}{\mu_i \, T_\odot(r)}}\right) \bigg \} ~.
\end{eqnarray}

\subsection{Velocity-dependent cross section: $v_{\rm rel}^2$}

In a similar way, for $\sigma_{i,v_{\rm rel}^2} \propto v_{\rm rel}^2$, the expression for the $\cos\theta$ integral reads
\begin{eqnarray}
\label{eq:DEv2}
\int_{-1}^1 \dd \cos\theta \,  |\boldsymbol{w} - \boldsymbol{u}|  \, \langle \Delta E_i \rangle & = & \frac{\mx}{140 \, \mu_{i,+}^2 \, v_0^2} \, \frac{1}{w \, u} \, \bigg[ \bigg(\mu_i \, (6 \, w + u) + (w + 6 \, u)\bigg) \, (w-u)^5 \, |w-u|  \nonumber \\
& & \hspace{2.7cm} - \bigg((6 \, \mu_i + 1) \, w - (\mu_i + 6)\bigg) \, (w + u)^6)\bigg] ~.
\end{eqnarray}
Again, the velocity integrals can be performed analytically. In the case with no cutoff in the DM velocity distribution, the equivalent equation to Eq.~(\ref{eq:tx-s}) is
\begin{equation}
\label{eq:tx-p}
\sum_i \int_0^{\rm{R_\odot}} \sigma_
{i,0} \,  \, n_i(r)\,  \frac{\mx m_i}{(m_i + \mx)^2} \, \left(\frac{m_i \tx + \mx T_\odot(r)}{\mx m_i}\right)^{3/2} \, \left( T_\odot(r) - \tx  \right) \, e^{\frac{- \mx \, \phi(r)}{ \tx}} \, r^2 \, \dd r = 0 ~.
\end{equation}
The expression for $\epsilon_{i,v_{\rm rel}^2}$ with cutoff at a velocity $v_c(r)$ reads
\begin{eqnarray}
\label{eq:epsT-v2}
\epsilon_{i,v_{\rm rel}^2}(r,\tx,T_c) & = & \frac{N_\chi \, e^{- \mx \,  \phi(r)/\tx}}{\left( \text{Erf}\left(\sqrt{\frac{T_c(r)}{\tx}}\right) - \sqrt{\frac{4 \, T_c(r)}{\pi \, \tx}} \, e^{-T_c(r)/\tx} \right) \, 4 \pi \int_0^{R_\odot} \dd r \, r^2  \, e^{-\mx \phi(r)/\tx}}  \nonumber \\ 
& & \times \frac{12}{\mx} \sqrt{\frac{2}{\pi \, \mx}} \, \frac{\mu_i}{\mu_{i,+}^2} \, n_i(r) \, \frac{\sigma_{i,0}}{v_0^2} \, \left\{\text{Erf}\left(\sqrt{\frac{T_c(r)}{\mu_i \, T_\odot(r)}}\right) \, \frac{e^{-T_c(r)/\tx}}{\sqrt{\tx}} \, \right. \nonumber \\
& & \times \, \left[ 3 \, \left( \frac{3 \, \mu_i^2}{4} T_\odot^2(r) - 2 \, T_c^2(r) - \left(2 \, \tx + \frac{3 \, \mu_i}{2} \, T_\odot(r)\right)^2 \right) \, \left(T_\odot(r) - \tx - T_c(r) \right) \right. \nonumber \\
& & \left. + \, \left( 9 \, \mu_i \, T_c(r) \, T_\odot(r) - 12 \, \tx \, T_\odot(r) - 18 \, \mu_i \, T_\odot^2(r) - 4 \, T_c^2(r) - \frac{3 \, \mu_i^3}{4} \, \frac{T_\odot^3(r)}{T_c(r)} \right) \, T_c(r)   \right] \nonumber \\
& & + \, \sqrt{\frac{\mu_i \, T_c(r) \, T_\odot(r)}{\pi \, \tx}} \, e^{-\left(\frac{T_c(r)}{\tx} + \frac{T_c(r)}{\mu_i T_\odot(r)}\right)}  \left[ \left(\frac{\mu_i \, \mx}{16} \frac{T_\odot(r)}{T_c(r)} + \frac{\mu_i}{30} \, T_\odot(r) + \frac{T_c(r)}{6} \right) \, T_c(r) \right. \nonumber \\
& & \left. - \, \left(\tx + \frac{5 \, \mu_i}{4} \, T_\odot(r) + \frac{T_c(r)}{2}\right) \, (T_\odot(r) - \tx) \right] \nonumber \\
& & \left. + \, \left(\tx + \mu_i \, T_\odot(r)\right)^{3/2} \, (T_\odot(r) - \tx) \, \text{Erf}\left(\sqrt{\frac{T_c(r)}{\tx} + \frac{T_c(r)}{\mu_i T_\odot(r)}}\right)
\right\}
\end{eqnarray}

\subsection{Momentum-dependent cross section: $q^2$}

For $q^2$-dependent cross sections, the results are the same as for the $v_{\rm rel}^2$-dependent case, except from a constant factor,
\begin{equation}
\label{eq:eps-q2}
\epsilon_{i, q^2} (r,\tx,T_c) = \frac{4}{3} \, \left(\frac{\mx \, v_0}{q_0}\right)^2 \, \epsilon_{i, v_{\rm rel}^2} (r,\tx,T_c) ~.
\end{equation}
Therefore, the temperature $\tx$ in the isothermal approximation for the $v_{\rm rel}^2$-dependent and $q^2$-dependent cases, with the cross sections as defined in this work, is the same as long as condition of no net energy flow, Eq.~(\ref{eq:nonetflow0}), is applied.

\subsection{Correction to the temperature calculation: including evaporation}

So far, as done always in the literature, to compute the temperature of the isothermal distribution in the optically thin regime, we have assumed there is not net flow of energy carried away by DM particles, i.e., DM particles are assumed to be confined within the Sun. However, for low DM masses (typically below a few GeV), evaporation from the Sun is very efficient, so there is indeed a net flux of energy. Indeed, a correction becomes crucial to even find a solution in some cases. Therefore, Eq.~(\ref{eq:nonetflow0}) needs to be modified in order to take into account the energy carried out of the Sun by the evaporated DM particles, i.e., 
\begin{equation}
\label{eq:newTeq}
\sum_i \int_0^{R_\odot} \epsilon_i(r, \tx, T_c) \, 4 \pi r^2 \, \dd r = \sum_i \int_0^{R_\odot} \epsilon_{{\rm evap}, i}(r, \tx, T_c) \, 4 \pi r^2 \, \dd r  ~,
\end{equation}
where, in general,
\begin{equation}
\label{eq:espevapgen}
\epsilon_{{\rm evap},i}(r,\tx,T_c) = \int \dd^3 \boldsymbol{w} \, n_{\chi,{\rm iso}}(r,t) \, f_{\chi, {\rm iso}}(\boldsymbol{w},r) \, \left[ \int_{v_e(r)}^{w} K_i^- (w \rightarrow v) \, \dd v + \int_{w}^{\infty} K_i^+ (w \rightarrow v) \, \dd v \right] ~, 
\end{equation}
with the rates $K_i^{\pm} (w \to v)$ being equivalent to $R_i^{\pm} (w \to v)$, but including $\Delta E_i$, i.e., 
\begin{eqnarray}
\label{eq:K}
K_i (w \to v) & = & \int n_i(r) \, \frac{\dd \sigma_i}{\dd v} \, |\boldsymbol{w} - \boldsymbol{u}| \, \Delta E_i \, f_i(\boldsymbol{u},r) \, \dd^3 \boldsymbol{u} \nonumber \\
& = &  \Delta E_i \, R_i(w \to v) = \frac{\mx}{2} \, \left(v^2 - w^2\right) \, R_i(w \to v) ~.
\end{eqnarray}
If there is a cutoff in the DM velocity distribution such that $w \le v_c(r) \le v_e(r)$, then Eq.~(\ref{eq:espevapgen}) gets simplified to
\begin{equation}
\label{eq:espevap}
\epsilon_{{\rm evap},i}(r,\tx,T_c) = \int_0^{v_c(r)} n_{\chi,{\rm iso}}(r,t) \, f_{\chi, {\rm iso}}(\boldsymbol{w},r) \, 4 \pi w^2 \, \dd w \, \int_{v_e(r)}^{\infty} K_i^+ (w \rightarrow v) \, \dd v   ~.
\end{equation}
Therefore, although the left-hand sides of Eq.~(\ref{eq:newTeq}) for $v_{\rm rel}^2$-dependent and $q^2$-dependent cross sections are proportional, the right-hand sides are not, which results in slightly different temperatures for low masses, for which the right-hand side matters. Also notice that we have not included the suppression factor appearing in the general expression of the evaporation rate, Eq.~(\ref{eq:ev+}), because the approximation of the isothermal distribution is only valid in the Knudsen limit, for which $s(r) = 1$.

\section{Propagation of dark matter in the Sun}
\label{app:prop}

\subsection{Mean free path}
\label{app:mfp}

The total mean free path of DM particles in the solar medium is defined as $\ell^{-1}(r) = \sum_i \ell_i^{-1}(r)$, where $\ell_i^{-1}(r) = \langle \sigma_i \rangle (r) \, n_i(r)$ is the partial mean free path for DM interactions with a thermal averaged scattering cross section $\langle \sigma_i \rangle (r)$ off targets with density $n_i(r)$. This thermal average is performed over the two initial  (DM and targets) velocity distributions, i.e.,
\begin{equation}
\label{eq:avgsigmaapp}
\ell_i^{-1}(r) = n_i(r) \, \langle \sigma_i \rangle (r) = n_i(r) \, \int \dd^3 \boldsymbol{w} \, f_{\chi}(\boldsymbol{w},r) \int \dd^3 \boldsymbol{u} \, f_i(\boldsymbol{u},r) \, \sigma(\boldsymbol{w},\boldsymbol{u}) ~.
\end{equation}
For velocity-independent and isotropic cross sections, with or without velocity cutoff, trivially,
\begin{equation}
\langle \sigma_i \rangle (r) = \sigma_{i, 0} ~.
\end{equation}
For the $v_{\rm rel}^2$-dependent cross sections considered in this work, $\sigma_{i,v_{\rm rel}^2} \propto v_{\rm rel}^2$, Eq.~(\ref{eq:dsvn}), the thermal average reads
\begin{eqnarray}
\label{eq:avgsigv2} 
\langle \sigma_{i,v_{\rm rel}^2} \rangle (r) & = & \frac{\sigma_{i,0}}{v_0^2} \, 6 \, \mu_{i,+} \,  \frac{T_\odot(r)}{\mx} \\ 
& & \times \left(\frac{\Sigma_{i, \rm LTE}(r) \, \mathfrak{f}(K) \, n_{\chi,{\rm LTE}}(r, t) + \Sigma_{i, \rm iso}(r) \, \left(\frac{\mu_i \, T_\odot(r) + \tx}{2 \, \mu_{i,+} \, T_\odot(r)}\right) \, \left(1-\mathfrak{f}(K)\right) \, n_{\chi,{\rm iso}}(r, t)}{\mathfrak{f}(K) \, n_{\chi,{\rm LTE}}(r, t) + \left(1-\mathfrak{f}(K)\right) \, n_{\chi,{\rm iso}}(r, t)}\right) ~, \nonumber
\end{eqnarray}
where the position-dependent normalizations are given by
\begin{equation}
\label{eq:normLTE}
\Sigma_{i, \rm LTE} (r) = 1 - \frac{1}{6 \, \sqrt{\pi} \, \mu_{i,+}} \, \left(\frac{v_c(r)}{{\tilde v}_\chi(r)}\right)^3 \, \frac{e^{- v_c^2(r)/{\tilde v}_\chi^2(r)}}{\textrm{Erf}\left(\frac{v_c(r)}{{\tilde v}_\chi(r)}\right) - \frac{2}{\sqrt{\pi}} \, \frac{v_c(r)}{{\tilde v}_\chi(r)} \,  e^{- v_c^2(r)/{\tilde v}_\chi^2(r)}}
\end{equation}
\begin{equation}
\label{eq:normiso}
\Sigma_{i, \rm iso} (r) = 1 - \frac{\mu_i \, T_\odot(r)}{3 \, \sqrt{\pi} \, \left(\mu_i \, T_\odot(r) + \tx\right)} \, \left(\frac{v_c(r)}{u_i(r)}\right)^2 \, \left(\frac{v_c(r)}{\vx}\right) \, \frac{e^{- v_c^2(r)/\vx^2}}{\textrm{Erf}\left(\frac{v_c(r)}{\vx}\right) - \frac{2}{\sqrt{\pi}} \, \frac{v_c(r)}{\vx} \,  e^{- v_c^2(r)/\vx^2}} ~, 
\end{equation}
with ${\tilde v}_\chi^2(r) \equiv 2 \, T_\odot(r)/\mx$ and $\vx^2 = 2 \, \tx/\mx$. In case of no cutoff, $\Sigma_{i, \rm LTE}(r) = \Sigma_{i, \rm iso}(r) =1$.

The calculation of the mean free path is mainly relevant in the conduction limit (optically thick regime), i.e., when $\tx(r) = T_\odot(r)$ and $ \mathfrak{f}(K) = 1$, in which Eq.~(\ref{eq:avgsigv2}) gets simplified as
\begin{equation}
\label{eq:avgsigLTE}
\langle \sigma_{i,v_{\rm rel}^2} \rangle (r) = \frac{\sigma_{i,0}}{v_0^2} \, 6 \, \mu_{i,+} \, \frac{T_\odot(r)}{\mx} \, \Sigma_{i, \rm LTE} ~,
\end{equation}
Similarly, for $q^2$-dependent cross sections, $\dd \sigma_{i,q^2}/\dd \cos\theta_{\rm cm} \propto q^2$, Eq.~(\ref{eq:dsqn}), 
\begin{equation}
\label{eq:avgsigq2}
\langle \sigma_{i,q^2} \rangle (r,v_c) = \left(\frac{\mx \, v_0}{q_0}\right)^2 \, \langle \sigma_{i,v_{\rm rel}^2} \rangle (r,v_c) ~,
\end{equation}
and thus, the mean free paths for $v_{\rm rel}^2$-dependent and $q^2$-dependent cross sections coincide, as long as $\sigma_{i,0}$ is the same and $q_0 = \mx \, v_0$, for the cross sections as defined here.

\subsection{Suppression factor}
\label{app:supp}

The suppression factor $s(r)$ that accounts for the fraction of DM particles with velocities larger than the escape velocity (after the first interaction) that escape the Sun was defined in Eq.~(\ref{eq:s}) as~\cite{Gould:1990}
\begin{equation}
\label{eq:sapp}
s(r) = \eta_{\rm ang}(r) \, \eta_{\rm mult}(r) \, e^{-\tau(r)} ~,
\end{equation}
where $\tau (r) = \int_r^{R_\odot} \ell^{-1} (r') \, \dd r'$ is the optical depth in the radial direction. The optical depth in a generic direction, denoted by the angle with respect to the radial direction, $\theta_r$, is given by
\begin{equation}
\label{eq:taugen}
\tau (r, \cos{\theta_r}) = \int_0^{l_\odot (r, \cos{\theta_r})} \ell^{-1}(r') \, \dd l ~,
\end{equation}
where $l_\odot (r, \cos{\theta_r}) = - r \cos{\theta_r} + \sqrt{R_\odot^2 - r^2 \, \sin^2 \theta_r}$ and $r' = \sqrt{l^2 + r^2 + 2 \, l \, r \, \cos{\theta_r}}$~.

The factor $\eta_{\rm ang} (r)$ takes into account that DM particles move in non-radial orbits and is defined as
\begin{equation}
\label{eq:etaangdef}
\eta_{\rm ang}(r) =   e^{\tau(r)} \, \int_{-1}^{1}  \gamma(\theta_r) \, e^{- \tau(r,\cos{\theta_r})} \, \dd \cos{\theta_r} ~,
\end{equation}
where $\gamma(\theta_r)$ is the DM distribution in the polar angle. This was estimated in Ref.~\cite{Gould:1990} assuming an isotropic distribution ($\gamma_0(\theta_r) = 1/2$) and including an approximate correction from the dipolar contribution ($\gamma_1 (\theta_r) \propto \cos{\theta_r}$). In order to allow for a smooth transition between the optically thin and thick regimes, the estimate of Ref.~\cite{Gould:1990} can be modified as~\cite{Bernal:2012qh}
\begin{equation}
\label{eq:etaang}
\eta_{\rm ang}(r) = \frac{7}{10} \, \frac{1 - e^{-10 \, \tau(r)/7}}{\tau(r)} ~.
\end{equation}
Note that for $\tau \lesssim 3$, this is a better estimate of $\eta_{\rm ang} (r)$ than that of Ref.~\cite{Gould:1990}. However, also note that this factor has extra dependences on $\mu_i$ and on the type of interaction, but we do not refine it any further here.

The factor $\eta_{\rm mult}(r)$ takes into account the possibility of DM particles escaping even after interacting several times. Following Ref.~\cite{Gould:1990}, we have estimated it in the general case when the probability of one DM particle with energy $E_\chi$ loosing a fraction of energy $\hat E_\chi (r) \equiv \frac{E_\chi}{T_\odot(r)} - \hat \phi(r)$ after one interaction, with $\hat \phi(r) \equiv \frac{\mx \, v_e^2(r)}{2 \, T_\odot(r)}$ the local dimensionless escape energy, is given by $1 - e^{-\hat E_\chi(r)/\hat \phi(r)}$, instead of simply $\hat E_\chi(r)/\hat \phi(r)$, the latter being valid when this probability is small. In this case~\cite{Gould:1990},
\begin{eqnarray}
\label{eq:etamult0}
\eta_{\rm mult} (r) & = & e^{\tau(r)} \, \sum_{n=0}^{\infty} \int_0^\infty \dd \hat E_\chi(r) \, \left(\frac{1}{n!} \left(1- e^{-\hat E_\chi(r)/\hat \phi(r)}\right)^n \right) \, \left( e^{- \tau(r)} \, \frac{\tau^n(r)}{n!} \right) \, e^{-\hat E_\chi(r)} \nonumber \\
& = & {}_{0}F_{1} \left( ; 1 + \hat \phi(r) ; \tau(r)\right) ~, 
\end{eqnarray}
where the first term in the integral represents the probability that the DM particle looses at least an energy $\hat E_\chi(r)$ after $n$ interactions, the second term is the probability of $n$ scatterings and the third term is the initial DM distribution above the escape velocity. The result is the confluent hypergeometric limit function ${}_{0}F_{1} \left( ; b ; z\right)$. In order to correct for the fact that the probability of scattering to higher energies is finite because the medium is not at zero temperature and to correct the collision rate to take into account the targets thermal velocity in the DM-target relative velocity, in analogy with Ref.~\cite{Gould:1990}, finally we estimate it as
\begin{equation}
\label{eq:etamult}
\eta_{\rm mult} (r) = {}_{0}F_{1} \left( ; 1 + \frac{2}{3} \, \hat \phi(r) ; \tau(r)\right) ~.
\end{equation}
In the limit $\hat \phi(r) \gg 1$, which applies to most of the relevant parameter space, this factor reduces to $\eta_{\rm mult} (r) = e^{1.5 \, \tau(r)/\hat \phi(r)}$~\cite{Gould:1990}. In the opposite limit, $\hat \phi(r) \ll 1$, $\eta_{\rm mult} (r) = I_0 \left(2 \, \sqrt{\tau(r)}\right)$, where $I_0(x)$ is the modified Bessel function of the first kind of order 0.
	
\small
	
\bibliographystyle{JHEP-mod}
\bibliography{biblio}	
	
\end{document}